\documentclass[10pt,twocolumn,twoside]{IEEEtran}
\usepackage{epsfig,color,amsmath}
\usepackage[noadjust]{cite}
\usepackage{amsthm}
\IEEEoverridecommandlockouts
\usepackage{wrapfig} 
\usepackage{bm}
\usepackage{epstopdf}
\usepackage{amssymb}
\usepackage{url}
\usepackage{enumitem}
\usepackage{multirow}
\usepackage{makecell}
\usepackage{algorithm,algorithmic}	
\setlist[itemize]{leftmargin=*}

\DeclareMathOperator*{\minimize}{minimize}

\DeclareMathOperator{\subjectto}{subject\ to}

\makeatother

\newtheorem{theorem}{Theorem}
\newtheorem{mydef}{Definition}
\newtheorem{mylem}{Lemma}

\newtheorem{rem}{Remark}
\newtheorem{asmp}{Assumption}

\newcommand{\mat}[1]{\boldsymbol{#1}}

\newcommand{\bmat}[1]{\begin{bmatrix} #1 \end{bmatrix}}

\providecommand{\mC}{\ensuremath{\mat{C}}}
\providecommand{\mD}{\ensuremath{\mat{D}}}

\providecommand{\mI}{\ensuremath{\mat{I}}}

\newcommand{\st}{{\rm s.t.}}

\newcommand{\m}{\boldsymbol}
\allowdisplaybreaks[4]
\newcommand{\mc}[1]{\mathcal{#1}}
\newcommand{\mbb}[1]{\mathbb{#1}}
\newcommand{\mr}[1]{\mathrm{#1}}

\usepackage{graphicx}
\usepackage{ifpdf}
\usepackage{epstopdf}
\usepackage{ifpdf}
\ifpdf
\DeclareGraphicsExtensions{.eps}
\else
\DeclareGraphicsExtensions{.eps}
\fi
\usepackage[caption=off]{subfig} 

\usepackage[colorlinks = true,
linkcolor = blue,
urlcolor  = blue,
citecolor = blue,
anchorcolor = blue]{hyperref}
\title{\textcolor{black}{Robust Control for Renewable-Integrated Power Networks Considering Input Bound Constraints and Worst-Case Uncertainty Measure}}
\author{Ahmad F. Tah$\text{a}^1$, Mohammadhafez Bazrafsha$\text{n}^1$,  Sebastian Nugroh$\text{o}^1$, Nikolaos Gatsi$\text{s}^1$, and Junjian Q$\text{i}^2$
\thanks{$^1$Department of Electrical and Computer Engineering, The University of Texas at San Antonio, TX 78256.
	$^2$Department of Electrical and Computer Engineering, University of Central Florida, Orlando, FL 32816.
	Emails: \{Sebastian.Nugroho, Mohammadhafez.Bazrafshan, Ahmad.Taha, Nikolaos.Gatsis\}@utsa.edu, Junjian.Qi@ucf.edu.
	This material is based upon work supported by the National Science Foundation under Grants CCF-1421583, CBET-1637249, and CMMI-1728629. }}
\begin{document}
\maketitle

\begin{abstract}
Uncertainty from renewable energy and loads is one of the major challenges for stable grid operation. Various approaches have been explored to remedy these uncertainties. In this paper, we design centralized or decentralized state-feedback controllers for generators while considering worst-case uncertainty. Specifically, this paper introduces the notion of $\mathcal{L}_{\infty}$ robust control and stability for uncertain power networks.  Uncertain and nonlinear differential algebraic equation model of the network is presented. The model includes unknown disturbances from renewables and loads. Given an operating point, the linearized state-space presentation is given. Then, the notion of $\mathcal{L}_{\infty}$ robust control and stability is discussed, resulting in a nonconvex optimization routine that yields a state feedback gain mitigating the impact of disturbances. The developed routine includes explicit input-bound constraints on generators' inputs and a measure of the worst-case disturbance. The feedback control architecture can be centralized, distributed, or decentralized. Algorithms based on successive convex approximations are then given to address the nonconvexity. Case studies are presented showcasing the performance of the $\mathcal{L}_{\infty}$ controllers in comparison with automatic generation control and $\mathcal{H}_{\infty}$ control methods.
\end{abstract}

\section{Introduction, Literature Review, Paper Contributions and Organization}
Wide area measurement systems, phasor measurement units and advanced communication technologies provide the needed assets to transform traditional power grids from hierarchical, unobservable systems to integrated, resilient ones. Specifically, advances in smart grids present major solutions to power grids' major challenges: robustness against uncertainty from intermittent renewable energy generation and loads. 

A plethora of research studies explore solutions to the aforementioned challenge. These solutions can be organized into four categories. The first category explores data mining and estimation methods to better predict wind speeds, solar irradiance, and loads~\cite{lei2009review}. This allows for improved planning, operation, and real-time control. Unfortunately, deviation in wind speed and solar irradiance is still significant, in comparison with models that predict demand. The second category studies the design of grid operating points for generators with lower operational costs and desirable stability properties~\cite{MalladaTang2013,bazrafshan2017coupling}.  The third category pertains to the design of economic incentives and demand-response methods that drive users to consume less energy,  thereby impacting the overall grid generation and the stability of the grid~\cite{siano2014demand}. The fourth category of research investigates the design of robust, real-time centralized/decentralized controllers for traditional generators or distributed energy resources ensuring that the grid is operating within its limits. These methods have been applied for different power system applications and contexts. This paper focuses on the fourth category of methods. We briefly review the most relevant literature, while acknowledging that the review given next is by no means exhaustive. 

A linear matrix inequality (LMI)-based criterion to \emph{assess} small-signal voltage stability in the presence of uncertain time constants of dynamic loads  is  given in \cite{Nguyen2016}.  Construction of LMI-based energy functions via convex approximations in \cite{Vu2017} provide guarantees convergence of post-fault dynamics to stable equilibrium points.   These works suggest remedial control actions but do not design controllers.  
Governor-based robust decentralized controller designs using LMIs have been initially pursued for transient-stability in \cite{Siljak2002}, later extended to primary frequency control in \cite{Marinovici2013}, and recently developed for wide-area control (WAC) in \cite{Lian2017}. In addition to decentralization, the major strengths of these works is that they avoid linearization around operating points. However, to derive controllers, networks with only generators (by considering model reduction) are considered and bounds on system nonlinearities are assumed---bounds that tend to be conservative \cite{Marinovici2013}.

The linear quadratic regulator (LQR), and more generally, $\mathcal{H}_2$ robust controllers for power systems have also been extensively researched. LQR  was initially used for secondary frequency control~\cite{Fosha1970}. More recently,  $\mc{H}_2$ controllers have found application in wide-area feedback controllers that are used in conjunction with power system stabilizers.  An efficient solution method for structured $\mc{H}_2$ problems from \cite{Lin2011} is employed to obtain optimal controllers that feed back a limited set of measurements to generator automatic voltage regulators~ \cite{Naguru2015}.  Sparsity-constrained LQR control for WAC is also considered in \cite{LianChakrabortty2017}, in which a decentralized solution algorithm is computed by casting the problem as non-cooperative game.

The property that weighted $l_1$ minimization yields sparse solutions \cite{CandesBoyd2008} has lead to the development of novel WAC methods.  For instance, the ADMM solution of $l_1$ augmented problems initially proposed in \cite{Lin2013} has been applied for WAC via voltage regulators in
\cite{Dorfler2014}, which requires slack-bus reference angle measurements, and in \cite{Wu2016a} that bypasses the aforementioned requirement. 
Similar WAC methods have been employed in \cite{PiroozAzad2016} to control rectifier current and inverter voltage setpoints of HVDC links.  A faster  algorithm for the same $l$-1 augmented formulation is developed in \cite{Wytock2013} using a proximal Netwon method and is applied to WAC. 

A decentralized discrete-time LQR controller is designed in \cite{Singh2016}  for synchronous generators where terminal voltage and angle measurements are treated as pseudo-inputs.  Using normal forms and by modeling network loads as constant impedances, \cite{Singh2017} decomposes the nonlinear DAEs of the power system into equivalent linearized and internal dynamics.  LQR then stabilizes the linearized dynamics while system internal dynamics are shown to be provably asymptotically stable irrespectively of operating conditions, further extending the application of LQR to cover both small-signal and transient stability.

Moving on from LQR and $\mc{H}_2$ controllers, $\mc{H}_{\infty}$ controllers have also been recently investigated. Sparse $l_1$ regularized $\mc{H}_{\infty}$  WAC controller design is obtained in \cite{Schuler2014a}  via an optimization problem with nonlinear matrix inequalities. The work of \cite{Schaab2017} presents in an LMI-based decentralized $\mc{H}_{\infty}$ controller for synchronous generators and doubly-fed induction generators that addresses network transient and voltage stability. Centralized $\mc{H}_{\infty}$ controllers robust to load and renewable disturbances are developed in \cite{Bevrani2016} to aid secondary frequency regulation in islanded microgrids, albeit simplistic first-order models for distributed generators are assumed. 

The approaches presented in \cite{Fosha1970,Naguru2015,LianChakrabortty2017,Dorfler2014,PiroozAzad2016,Wytock2013,Singh2016,Singh2017,Schuler2014a,Schaab2017,Bevrani2016} do not explicitly consider bounds on the controllable input of generators. Practical equipment considerations on the other hand may necessitate bounds on the instantaneous actuation effort. Also, the variation of constant-power renewables is not explicitly leveraged. Furthermore, even though constraints on the feedback gain are considered in previous works \cite{Siljak2002,Marinovici2013,Lian2017}, a network-reduced model of only generator buses is utilized. The direct impact of constant-power nonsynchronous renewable generators and loads is abstracted in bounds that tend to become too conservative.

This paper introduces the notion of $\mathcal{L}_\infty$ control---originally proposed in~\cite{pancake2000d} for generic dynamic systems and expanded in this paper---to power systems with high uncertainty from renewable energy and loads. This notion of $\mathcal{L}_\infty$ control is different than $\mathcal{H}_{\infty}$ control.
The $\mc{H}_{\infty}$ norm of a linear system with transfer function $\m G$ under disturbances $\m w(t)$ is the worst-case, induced energy-to-energy gain of the closed-loop system. On the other hand, $\mc{L}_{\infty}$ control is concerned with the $\mc{L}_{\infty}$ gain of the system, that is, the gain of the system when viewed as an operator acting on $\mc{L}_{\infty}$ inputs and producing $\mc{L}_{\infty}$ outputs. With the aforementioned differences in mind, $\mc{L}_{\infty}$-stability is a concept that we introduce for power networks in this paper.
Both the $\mc{L}_{\infty}$ and $\mc{H}_{\infty}$ control problems yield state feedback gains that remedy the impact of disturbances. The paper contributions are as follows.
\begin{itemize}
	\item From a control-theoretic perspective, the methods presented in this paper advance the concept of $\mc{L}_{\infty}$ control to incorporate (a) strict control input constraints, (b) an explicit measure of the worst-case unknown disturbance into the controller design, (c) formulation of a nonconvex optimization routine that seeks to obtain an optimal (or locally optimal) solution to the main $\mc{L}_{\infty}$ control problem, in comparison with the iterative algorithm in~\cite{pancake2000d} that only yields a feasible solution, and (d) derivation of successive convex approximation algorithms---with convergence guarantees---to solve the nonconvex $\mc{L}_{\infty}$ problem yielding locally optimal solutions. All of these contributions are showcased for centralized feedback control, as well as distributed or purely decentralized control architectures. 
	\item From a power  network perspective, a fairly general differential algebraic model of the power grid is considered. This model encapsulates the algebraic power flow and stator equations as well as fourth-order generator dynamics with simplified governor and exciter control inputs. Based on linearization around a known equilibrium, a linear (centralized and decentralized) state-feedback controller is computed that ensures a given performance metric is met while the adverse effects of uncertainty from wind, solar, and load prediction errors are mitigated.  The magnitude of the worst-case disturbance (which can be given by a system operator a day ahead) and the grid's operating point are thus leveraged to compute generator control actions.  The resulting controller finds applications in secondary frequency or wide-area control of power systems. 
	\item The performance of the centralized and decentralized $\mc{L}_{\infty}$ controllers when applied to the high-order nonlinear grid model---amidst significant prediction errors from wind and solar generation---illustrates that the controllers are able to stabilize the nonlinear power network model. A comparison of the proposed control algorithm with automatic generation control (AGC) and $\mc{H}_{\infty}$ control is also presented. 
\end{itemize}
Section~\ref{sec:syncmodel} presents the uncertain power network model. Section~\ref{sec:robustcontrol} develops the worst-case, robust $\mathcal{L}_{\infty}$ controller for the network model, yielding a nonconvex optimization routine for a centralized architecture with full state feedback. Section~\ref{sec:BMIs} explores successive convex approximations for the nonconvex problem, and Section~\ref{sec:decentralized} develops the $\mc{L}_{\infty}$ controller for decentralized control architectures. Finally, numerical tests conclude the paper in Section~\ref{sec:numtests}. The paper's notation is given next.

\textcolor{black}{The symbols $\mathbb{R}^n$ and $\mathbb{R}^{p\times q}$ denote column vectors with $n$ elements and real-valued matrices with size $p$-by-$q$. The set of $n\times n$ symmetric and positive definite matrices are denoted by $\mathbb{S}^{n}$ and $\mathbb{S}^{n}_{++}$. Italicized, boldface upper and lower case characters represent matrices and column vectors---$a$ is a scalar, $\m a$ is a vector, and $\m A$ is a matrix. Matrix $\m I_n$ is a $n\times n$ identity square matrix, while $\m 0$ and $\m O$ represent zero vectors and matrices of appropriate dimensions. The symbol `$\star$' is used to represent symmetric components in symmetric block matrices.}

\section{Renewable-Integrated DAE Network Model }~\label{sec:syncmodel}
We consider a power network with $N$ buses, modeled by a graph $(\mathcal{N},\mathcal{E})$, where $\mathcal{N}= \{1,\ldots, N\}$ is the set of nodes and $\mathcal{E} \subseteq \mathcal{N}\times\mathcal{N}$ is the set of edges; $\mathcal{N}_{i}$ denotes the neighboring nodes to node $i$. \textcolor{black}{Define the partition $\mathcal{N} = \mathcal{G} \cup \mathcal{L}$ where $\mathcal{G} = \{1,\ldots, G\}$ collects the buses containing $G$ synchronous generators and $\mathcal{L} = \{1,\ldots, L\}$ collects the $L$ buses that contain load buses only.   Further,  denote by $\mathcal{R} \subseteq \mc{N}$ the set of buses containing a total of $R$ renewable energy producers, such as solar and wind farms.}  Denote by $\m {a}_i(t)$ the vector of algebraic variables for all nodes $i \in \mathcal{N}$. For load nodes  $i \in \mathcal {L}$, there are  two algebraic variables, that is, $\m {a}_i(t)=\{v_i(t), \theta_i(t)\}$, where $v_i(t)$ and $\theta_i(t)$ denote the terminal load voltage and phase angle. 
\textcolor{black}{For generator nodes  $i \in \mathcal {G}$}, there are four  algebraic variables, that is, $\m {a}_i(t)=\{p_{g_i}(t), q_{g_i}(t), v_i(t), \theta_i(t)\}$, where  $p_{g_i}(t)$, $q_{g_i}(t)$,  $v_i(t)$, and $\theta_i(t)$ respectively denote generator real and reactive power, terminal voltage and phase angle. 
Italicized, boldface upper and lower case characters represent matrices and column vectors---$a$ is a scalar, $\m a$ is a vector, and $\m A$ is a matrix. Matrix $\m I$ is the identity square matrix, $\m 0$ and $\m O$ represent zero vectors and matrices of appropriate dimensions.
\subsection{Synchronous Generator Model}
We leverage the fourth order dynamics of synchronous generators with internal algebraic variables. The dynamics of synchronous generator $i \in \mathcal{G}$ can be written as~\cite{sauerpai1998}:
\begin{subequations} \label{SynGen}
		\begin{align}
			\dot{\delta_{i}} &= \omega_{i} - \omega_{s} \label{SynGen1} \\ 
			\dot{\omega_{i}} &= \frac{1}{M_{i}}\left[m_{i}-D_{i}(\omega_{i}-\omega_{s})-p_{g_{i}}\right] \label{SynGen2}    \\ 
			\dot{e_{i}} &= \frac{1}{\tau_{i}} \left[-\frac{x_{di}}{x'_{di}}e_{i} + \frac{x_{di} - x'_{di}}{x_{di}^{'}}v_{i}\cos(\delta_{i} - \theta_{i})+ f_{i} \right] \label{SynGen3} \\
			\dot{m_{i}} &= \frac{1}{T_{Ch_{i}}}\left[r_{i} - m_{i} - \frac{1}{R_i}(\omega_{i} - \omega_{s})\right], \label{SynGen4}  
		\end{align} 
\end{subequations}
where $\delta_{i}:=\delta_i(t)$, $\omega_{i}:=\omega_i(t)$, $e_{i}:=e_i(t)$, $m_{i}:=m_i(t)$, $r_{i}:=r_{i}(t)$, $f_{i}:=f_i(t)$ denote the generator rotor angle, rotor speed, internal electromotive force, mechanical input power, governor reference signal, and internal field voltage. $M_i$ is the rotor's inertia constant ($\mr{pu} \times \mr{s}^2$), $D_i$ is the damping coefficient ($\mr{pu} \times \mr{s}$), $\tau_{\mr{d}_i}$ is the direct-axis open-circuit time constant ($\mr{s}$), $x_{d_i}$ is the direct- axis synchronous reactance, $x'_{di}$ is the direct-axis transient reactance ($\mr{pu}$), $T_{Ch_{i}}$ and $T_{C_{i}}$ are the chest valve and reference valve time constants ($\mr{s}$),  $R$ defines the regulation constant of the speed-governing mechanism, and $\omega_{s}$ denotes the synchronous speed of rotor. In this work, we do not consider frequency-sensitive loads. If frequency-sensitive loads are placed on generator buses, their dynamics can be included in~\eqref{SynGen2}  by adjusting the coefficient $D_i$. 

Each synchronous generator has a total of four states, defined by $\m x_{s_{i}}(t) =[\delta_{i} \; \omega_{i} \; e_{i} \;  m_{i} ]^{\top}$, two control inputs, defined by $\m u_{s_{i}}(t) = [r_{i} \; f_{i}]^{\top} $, and four algebraic variables $\m a_{{s}_{i}}(t)= [ p_{g_{i}}\; q_{g_{i}} \; v_{i}\; \theta_{i}\; ]^{\top}.$ The following algebraic equations relate the generator real and reactive power output with generator voltage, internal EMF, and internal angle, and must hold at any time instant for generator nodes $i \in \mathcal{G}$~\cite{sauerpai1998}:
\begin{subequations} \label{SynGenPF}
	\begin{align}
		\begin{split}  \label{SynGenPF1}
			p_{g_{i}} ={}& \frac{e_{i}v_{i}}{x'_{di}}\sin(\delta_{i}-\theta_{i}) + \frac{x'_{di}-x_{qi}}{2x'_{di}x_{qi}}v_{i}^{2}\sin[2(\delta_{i}-\theta_{i})]   
		\end{split}  \\
		\begin{split}  \label{SynGenPF2}
			q_{g_{i}} ={}& \frac{e_{i}v_{i}}{x'_{di}}\cos(\delta_{i}-\theta_{i}) - \frac{x_{di}^{'}+x_{qi}}{2x_{di}^{'}x_{qi}}v_{i}^{2} \\ & + \frac{x_{di}^{'} -x_{qi}}{2x_{di}^{'}x_{qi}}v_{i}^{2}\cos[ 2(\delta_{i}-\theta_{i})] .
		\end{split}  
	\end{align}
\end{subequations}
In this paper, we focus on the small-signal stability of uncertain power systems.   By linearizing \eqref{SynGen} and \eqref{SynGenPF} around the operating point  $\{\m x_{s_{i}}^0, \m u_{s_{i}}^0, \m a_{s_{i}}^0\}$, we obtain the following dynamics of the small-signal system, with $\Delta \m x_s  := \m x(t) - \m x^0$
\begin{eqnarray}  \label{LSynGen1}
	\Delta \dot{\m x}_{{s}_{i}} = \m A_{s_{i}}  \Delta \m x_{{s}_{i}} + \m B_{s_{i}}  \Delta \m u_{{s}_{i}} + \m D_{s_{i}}  \Delta \m a_{{s}_{i}}    \label{LSynGen1b}
\end{eqnarray}
where $\m A_{s_{i}} , \m B_{s_{i}} , \m D_{s_{i}} $ are the Jacobian matrices corresponding to the linearization of the dynamics of synchronous generator $i$ around the operating point $\{\m x_{s_{i}}^0, \m u_{s_{i}}^0, \m a_{s_{i}}^0\}$. Similarly, \eqref{SynGenPF} can be linearized around $\{\m x_{s_{i}}^0, \m u_{s_{i}}^0, \m a_{s_{i}}^0\}$ as follows
\begin{eqnarray} \label{LSynGenPFG}
	\m 0 = \m H_{s_{x_{i}}}  \Delta \m x_{{s}_{i}}  + \m H_{s_{a_{i}}}\Delta \m a_{{s}_{i}} . \label{LSynGenPF1G}
\end{eqnarray}
The above dynamics \eqref{LSynGen1b} and \eqref{LSynGenPF1G} correspond to a single synchronous generator. Let $\m { \Delta x} =[\Delta  \m x_{s_{1}}^{\top} \;    \ldots \; \Delta  \m x_{s_{G}}^{\top} \; ]^{\top} $, $\m { \Delta u} = [\Delta  \m u_{s_{1}}^{\top} \;    \ldots \; \Delta  \m u_{s_{G}}^{\top} \; ]^{\top} $, $\m { \Delta a} = [\Delta  \m a_{s_{1}}^{\top} \;    \ldots \; \Delta  \m a_{s_{G}}^{\top} \; ]^{\top}$ define the states, control inputs, and algebraic variables for the $G$ synchronous generators in the power network. Given that, we obtain
\begin{subequations}  \label{GLSynGen}
	\begin{eqnarray} 
		\m {\Delta \dot{x}} &=&\m {A_{s}\Delta x} + \m {B_{s}\Delta u} + \m {B_{a}\Delta a}  \label{GLSynGena}        \\
		\m {0} &=& \m {H_{s_{x}}\Delta x}+ \m {H_{s_{a}}\Delta a} ,   \label{GLSynGenb}
	\end{eqnarray}
\end{subequations}
where $\m {A_{s}} \in \mathbb{R}^{4G\times 4G}$, $\m {B_{s}} \in \mathbb{R}^{4G\times 2G}$, $\m {D_{s}} \in \mathbb{R}^{4G\times 4G}$,  $\m {H_{s_{x}}} \in \mathbb{R}^{2G\times 4G}$.
For brevity, we do not provide the closed form presentation of these matrices. 
\subsection{Generation from Utility-Scale Solar and Wind Farms}~\label{sec:windd}
\color{black} 
Since the objective of this work is to obtain worst-case disturbance rejection controllers for the synchronous generators, we consider that predicted values of electric power generation from wind and solar buses $i \in \mc{R}$ are provided---similar to the widely available load forecasts.
Unlike traditional demand that can be predicted in hour-ahead markets within an accuracy of 1--5\% (see California ISO's daily hour-ahead prediction and actual demand~\cite{CAISO}), high-fidelity estimates of generation from wind and solar farms are difficult to obtain in day-ahead or hour-ahead fashion. Hence, we consider that real-time disturbances from $p_{r_{i}}(t)$ for buses $i \in \mc{R}$ that are unknown for the controller design. \normalcolor \textcolor{black}{Section~\ref{sec:numtests} includes concrete discussion on the choice of these unknown disturbances, as well as case studies demonstrating the performance of the $\mathcal{L}_{\infty}$ controller amidst large, unpredictable variations in generation from renewables.}
\subsection{Power Flow Equations and DAE Model with Uncertainty}~\label{sec:powerflow}
\textcolor{black}{For bus $i \in \mathcal{G} \cap \mc{R}$, the power flow equations of the power network can be written as}
{\begin{subequations} \label{GPF}
		\begin{align}
			\begin{split}   \label{GPF1} 
				p_{r_i}-p_{l_{i}} ={}& -p_{g_{i}} + G_{ii} v_{i}^{2} +\sum_{j = 1}^{N} (G_{ij} v_{i} v_{j} \cos(\theta_{ij})  \\
				&+ B_{ij}v_{i} v_{j}\sin(\theta_{ij})),\, 
			\end{split} \\
			\begin{split}   \label{GPF2}
				q_{r_i}-q_{l_{i}}={}& -q_{g_{i}} - B_{ii} v_{i}^{2} +\sum_{j = 1}^{N} (G_{ij} v_{i} v_{j} \sin(\theta_{ij}) \\
				& -B_{ij}v_{i} v_{j}\cos(\theta_{ij})),
			\end{split} 
		\end{align}
\end{subequations}}
\textcolor{black}{and for a bus $i \in \mathcal{L} \cap \mc{R}$, the power flow equations are}
{ \begin{subequations} \label{GPFe}
		\begin{align}
			\begin{split}  \label{GPF3}
				p_{r_{i}} -p_{l_{i}} ={}&  G_{ii} v_{i}^{2} +\sum_{j = 1}^{N} (G_{ij} v_{i} v_{j} \cos(\theta_{ij}) \\
				& + B_{ij}v_{i} v_{j}\sin(\theta_{ij})),\, 
			\end{split} \\
			\begin{split}  \label{GPF4}
				q_{r_{i}} -q_{l_{i}} ={}&  - B_{ii} v_{i}^{2} +\sum_{j = 1}^{N} (G_{ij} v_{i} v_{j} \sin(\theta_{ij})  \\
				&- B_{ij}v_{i} v_{j}\cos(\theta_{ij})),
			\end{split}
		\end{align}
\end{subequations}}where $\theta_{ij} = \theta_{i} - \theta_{j}$, $p_{l_{i}} = p_{l_{i}}(t)$, $q_{l_{i}} = q_{l_{i}}(t)$,  are the real and reactive power loads at bus $i$ modeled as time varying power load, and $p_{r_{i}} = p_{r_{i}}(t)$, $q_{r_{i}} = q_{r_{i}}(t)$ are the active and reactive power generated from the renewable energy sources at node $i$. 
Linearizing the power flow equation \eqref{GPF} and \eqref{GPFe} of all buses, we obtain the following relationship between the voltages, phase angles, active and reactive power
\begin{equation}~\label{LPFX}
 \hspace{-0.5cm}		\textcolor{black} {\bmat{ \Delta \m p_{r}-\Delta \m p_{l} \\ \Delta \m q_{r} -\Delta \m q_{l} }}
 = \underbrace{\begin{bmatrix}
			-\m I_{G} & \m O &\m  D_{{G}_{1}} & \m D_{{G}_{2}} \\ \m O& -\m I_{G} & \m D_{{G}_{3}} & \m D_{{G}_{4}} \\ \m O & \m O  & \m D_{{L}_{1}} & \m D_{{L}_{2}} \\ \m O & \m O & \m D_{{L}_{3}} & \m D_{{L}_{4}}
	\end{bmatrix}}_{\let\scriptstyle\textstyle
		\substack{ \m \Psi}} \begin{bmatrix}
		\Delta \m p_{g} \\ \Delta \m q_{g} \\ \Delta \m v \\ \Delta \m \theta
	\end{bmatrix},
\end{equation}
\textcolor{black}{where $\m \Psi \in \mbb{R}^{(2G+2L)\times (4G+2L)}$ is obtained by differentiating equations~ \eqref{GPF} and \eqref{GPFe} and obtaining the power flow Jacobian around the operating point of the power network. Specifically, $\m \Psi$ can be analytically obtained in terms of the power network parameters, but  for brevity, we do not include the exact structure of $\m \Psi$ as this would require lengthy listing of the closed form partial derivatives. }
We now define the state, controllable inputs, unknown inputs, disturbances, and algebraic variables of the uncertain power network. 
Combining the linearized power flow~\eqref{LPFX} with the internal algebraic equations of the synchronous generators~\eqref{GLSynGenb} as well as their associated dynamics \eqref{GLSynGena}, we obtain the following DAEs that model the dynamics of the uncertain network 
\begin{subequations}\label{DAE}
	\begin{align} 
		\m {\Delta \dot{x}}(t) &=\m {A_{s}\Delta x} (t)+ \m {B_{s}\Delta u}(t) + \m {B_{a}\Delta a}(t) \label{DAE1}\\
		\m {\Delta w}(t) &= \m H_x  \m {\Delta x}(t) + \m H_ u  \m {\Delta u} (t)+ \m H_a \m {\Delta a} (t), \label{DAE2}
	\end{align} 
\end{subequations}
where \textcolor{black}{$\Delta \m w= \begin{bmatrix}
(\Delta \m p_{r}^{\top}-\Delta \m p_{l}^{\top}) \; (\Delta \m q_{r}^{\top} -\Delta \m q_{l}^{\top})
\end{bmatrix}^{\top}\in \mbb{R}^{2G+2L}$} includes load and renewable energy deviations from the predicted values; matrices $\m H_x, \m H_u,$ and $\m H_a$ are all matrices of appropriate dimensions that include the linearization of the power network dynamics. 
\begin{asmp}
	Matrix $\m H_a$ is invertible. This assumption is mild as it holds for practical networks and for
	various operating points; see~\cite{MalladaTang2013} and references therein.
\end{asmp}
Assuming the invertibility of $\m H_a$, we can write
$$\Delta \m a(t) =  \m H_a^{-1} \left(\m {\Delta w}(t) - \m H_x  \m {\Delta x}(t)- \m H_ u  \m {\Delta u}(t)\right).$$ 
The DAEs in~\eqref{DAE} can then be written as
\begin{eqnarray}\label{Dynamics2}
	\m {{\Delta \dot x}} (t)= {\m A} \m {\Delta x} (t)+ {\m B}_u  \m {\Delta u} (t)+ \m B_w \m {\Delta w}(t) ,
\end{eqnarray}
where ${\m A} = {\m A}_s - \m B_a  \m H_a^{-1} \m H_x, {\m B}_u = {\m B}_s - \m B_a  \m H_a^{-1} \m H_u,
\m B_w=\m B_a \m H_a^{-1}$
In the next section, we discuss a robust control formulation that considers the worst case unknown inputs/disturbances $\Delta \m w(t)$ (from the uncertainty due to mismatch/deviation in load predictions and renewable energy generation) to obtain a state-feedback controller that drives the system to a neighborhood of the operating point. 
\section{Robust Feedback Control of Uncertain Power Networks}~\label{sec:robustcontrol}
Here, we present the $\mathcal{L}_{\infty}$ control formulation for the uncertain dynamics of the power network~\eqref{Dynamics2}. The objective of this formulation is to obtain a control law for the inputs of the synchronous generators $p_{\mathrm{ref}_{i}}$ and $f_{i}$, given the aforementioned disturbances. 
\subsection{Assumptions, Definitions, and Preliminaries}
\textcolor{black}{For the ease of exposition, we define
$n_x=4G, \; n_u=2G,$ and  $n_w=4G+2L$.} In summary, the uncertain system in~\eqref{Dynamics2} has $n_x$ states, $n_u$ controllable inputs, and $n_w$ unknown inputs. We also drop the $\Delta$ from the states, inputs, and disturbances, that is $\Delta \m x(t) \equiv \m x(t)$.
We now present the following needed assumptions and definitions.

\color{black}
\begin{mydef}
	The $\mathcal{L}_{\infty}$ space is defined as the set of signals which have bounded amplitude, that is
	\begin{align*}
		\mathcal{L}_{\infty} = \lbrace \m w:\mathbb{R}_+\rightarrow\mathbb{R}^{n_w} \,\vert \,\mathrm{sup}_{t\geq 0}\,\Vert \m w(t) \Vert_2 < \infty\rbrace,
	\end{align*} 
	and	the $\mc{L}_{\infty}$-norm of a signal $\m w \in \mc{L}_{\infty}$, denoted as $\Vert \m w \Vert_{\mc{L}_\infty}$, is given by
	$	\Vert \m w \Vert_{\mc{L}_\infty} = \underset{t \geq 0}{\mathrm{sup}} \, \Vert \m w(t) \Vert_2$ and $\Vert \m w \Vert_{2}$ is the 2-norm of $\m w$.
\end{mydef} \normalcolor 
This norm defines the worst-case value that the signal can take for $t\geq 0$. This implies that $\Vert \m w \Vert_{\mc{L}_\infty} \geq \Vert \m w(t) \Vert_2.$
\begin{asmp}~\label{ass2}
	The disturbance vector $\m w(t)$ belongs to the $\mc{L}_{\infty}$ space, and is considered to be completely unknown. 
\end{asmp}
\noindent \textbf{Design Requirement 1.}	\textit{A budget requirement $\vert\vert \m u(t) \vert\vert_2 \leq u_{\max}$ on the input $\m u(t)$ is given.}

\color{black}
Let $\m z(t) = \m C \m x(t) + \m D \m u(t)$ define the performance output of the control law of the power network which can include the deviations in the frequencies of the buses, as well as any other state of the synchronous generators. The performance index can also include the magnitude of the control actions which are essentially the deviations from the setpoints.

\begin{rem}
Individual performance indices $z_i(t)$ can be defined separately, as one performance index might focus only on the magnitude of the control action and another performance index can quantify frequency deviation of the most important bus in the network. In addition, and similar to the vintage LQR cost function $\m x^{\top}\m Q\m x + \m u^{\top}\m R\m u$, matrices $\m C$ and $\m D$ can be obtained from $\m Q$ and $\m R$ through the operator's preference of penalizing frequency and rotor angle deviations or penalizing higher magnitude of generators' control actions. Examples are given in Section~\ref{sec:numtests}. 
\end{rem}

\normalcolor 

Assumption~\ref{ass2} and Requirement 1 are practical, as the load disturbances and deviations in the wind speed and solar irradiance are naturally unknown inputs with bounded amplitudes, and the input budget for all the controls cannot exceed a certain predetermined limit. Next, we rewrite the dynamics augmented by the performance index
\begin{subequations}\label{Dynamics4}
	\begin{align}
		\m {{\dot x}}(t) &={\m A} \m { x}(t)+ {\m B}_u  \m { u}(t) + \m B_w \m {  w}(t)\\
		\m z(t)& =  \m C \m x(t) + \m D \m u(t).
	\end{align}
\end{subequations}
The objective of this section is to derive a control law $\m u(t)=\m K \m x(t)$ that minimizes the impact of the unknown inputs $\m w(t)$ on the performance index $\m z(t)$, while guaranteeing that the controller drives the system states to a neighborhood of the operating point. Given the feedback control law, the closed loop dynamics can be written as
\begin{align}~\label{equ:closedloop}
	\dot{\m x}(t) &= \m f(\m x, \m w) = (\m A+\m B_u\m K)\m x(t)+\m B_w \m w(t)\\
	{\m z}(t) &= \m h(\m x) =   (\m C+\m D\m K)\m x(t).
\end{align}
The next definition from \cite{pancake2000d} presents the properties of a special kind of robust dynamic stability, namely the $\mc{L}_{\infty}$ stability with performance level $\mu$.
\begin{mydef}~\label{def:LINF}
	The closed-loop system with unknown inputs~\eqref{equ:closedloop} is $\mathcal{L}_{\infty}$-stable with performance level $\mu$ if the following conditions are satisfied.
	\begin{enumerate}
		\item The closed-loop linear system without unknown inputs $\dot{\m x}(t)=\m f(\m x, \m 0)$ is asymptotically stable.
		\item For any unknown input $\m w(t) \neq \m 0$ and zero state initial conditions $(\m x_0=\m 0)$, we have
		$\Vert \m z(t) \Vert_2 \leq \mu \Vert \m w \Vert_{\mc{L}_\infty}$.
		\item For any nonzero initial conditions and unknown input, there exists a function $\beta:\mbb{R}^{n_x} \times \mbb{R}_{+} \rightarrow \mbb{R}_{+}$, such that
		$$ \small \hspace{-0.6cm}\Vert \m z(t) \Vert_2 \leq \beta \left(\m x_0,  \Vert \m w \Vert_{\mc{L}_\infty}\right) , \lim_{t \rightarrow \infty} \sup \Vert \m z(t) \Vert_2 \leq \mu \Vert \m w \Vert_{\mc{L}_\infty}.$$
	\end{enumerate}
\end{mydef} 
\subsection{$\mc{L}_{\infty}$ Controller Design with Input Bound Constraints}
The following theorem presents the design of the $\mc{L}_{\infty}$ state-feedback controller that aims to minimize the impact of the unknown disturbances on the state-performance. 
\begin{theorem}\label{th1}
	For the system defined in~\eqref{Dynamics4}, consider that the initial state value is $\m x_0$. Then, if there exist matrices $\m S=\m S^{\top}\succ 0$ and $\m Z$ and positive scalars $\{\alpha, \mu_0, \mu_1, \mu_2 \}$ that are the solution to the nonconvex optimization problem
	\begin{subequations}~\label{equ:LINF}
			\begin{align}
				f^* = \min& \;\;\;\mu_0 \mu_1 + \mu_2 \\
				\st&	\begin{bmatrix}
					\m S\m A^{\top}+\m A\m S\\ +\m Z^{\top}\m B_u^{\top}+\m B_u\m Z+\alpha \m S & \m B_w \\
				\star& -\alpha \mu_0 \m  I
				\end{bmatrix} \preceq 0 ~\label{equ:LINF-BMI1}\\
				& \begin{bmatrix}
					-\mu_1\m S & \m O & \m S\m C^{\top}+\m Z^{\top}\m D^{\top}\\  \star & -\mu_2\m I & \m O \\ \star& \star & -\m I
				\end{bmatrix}
				\preceq 0 ~\label{equ:LINF-BMI2}\\
				&\bmat{-\mu_0\rho^2 & \m x_0 \\ \star & -\m S} \preceq 0 \label{equ:LINF-BMI3}\\
				& \bmat{-\frac{u_{\max}^2}{\rho^2}\m S & \mu_0\m Z\\\star & -\mu_0 \m I} \preceq 0\label{equ:LINF-BMI4},
			\end{align}
	\end{subequations}
	then the feedback controller $\m u(t)=\m K \m x(t)$ with $\m K = \m Z\m S^{-1}$ guarantees that 
	$$ \Vert \m z(t) \Vert_2 \leq \mu \rho,\;\; \mu = \sqrt{\mu_0\mu_1+\mu_2} , \;\;  $$
where $\rho = \Vert \m w \Vert_{\mc{L}_\infty}$, and that the closed loop system with unknown inputs~\eqref{equ:closedloop} is $\mathcal{L}_{\infty}$-stable with performance level $\mu$. Furthermore, Design Requirement 1 is satisfied.
\end{theorem}

\textcolor{black}{The result in Theorem~\ref{th1} guarantees that the small-signal deviation in the performance index, that is $||\m z(t)||_2$, does not exceed the worst-case scenario of load and wind speed deviations from the setpoints (defined as $\mu ||\m w||_{\mc{L}_{\infty}})$, while satisfying the bound constraints on the control inputs. Specifically, this feedback $\mathcal{L}_{\infty}$ controller guarantees that $\m z(t)$ is in a tube of radius $\sqrt{\mu}  ||\m w||_{\mc{L}_{\infty}}=\sqrt{\mu} \rho$ of the operating point of the power network where \textit{(i)} $\mu$ is comprised of scalar optimization variables in~\eqref{equ:LINF} and \textit{(ii)} $\rho$ is a user-specified constant modeling worst-case uncertainty that contributes to the controller synthesis through the third and fourth  matrix inequality in~\eqref{equ:LINF}. In addition and in comparison with the results in~\cite{pancake2000d} which develop the $\mathcal{L}_{\infty}$ for general dynamic systems, Theorem~\ref{th1} includes \textit{(i)} the input bound constraints, \textit{(ii)} the explicit measure of the worst-case unknown input $\rho$, and \textit{(iii)} a nonconvex optimization routine that seeks to obtain an optimal (or locally optimal) solution to~\eqref{equ:LINF}.  In comparison, the iterative algorithm in~\cite{pancake2000d} only seeks a feasible solution.}
\color{black}
\begin{rem}
	The $\mc{L}_{\infty}$ controller from Theorem~\ref{th1} is robust to any bounded disturbance $\m w(t)$ defined earlier, granted a solution to~\eqref{equ:LINF} exists. With that in mind, the controller is not robust to changes in state-space matrices generated for various operating points. Robustness to changes in the operating point can be included through considering a polytopic version of the linearized dynamics and state-space matrices modeling various operating points.
	\end{rem}
\normalcolor

The next section is dedicated to solving the nonconvex problem~\eqref{equ:LINF} using convex optimization techniques.

\section{Successive Convex Approximations for~\eqref{equ:LINF}}~\label{sec:BMIs}
The nonconvex optimization problem~\eqref{equ:LINF} includes bilinear matrix inequalities (BMI) due to the presence of the terms $\alpha \m S$, $\alpha\mu_0 \m I$, and $\mu_1\m S$ in the first two constraints. It is very common in the robust control literature to introduce an alternating minimization-based algorithm to solve robust control problems with similar structure to~\eqref{equ:LINF}; see~\cite{pancake2000d}. However, these approaches do not typically provide optimality guarantees. In this section, we present a simple approach to solve~\eqref{equ:LINF} with convergence guarantees. The approach is based on expanding the BMIs as a difference of two convex functions that are then approximated by linear matrix inequalities (LMI).


To approximate BMIs with LMIs, we adopt the successive convex approximation (SCA) method that is introduced in \cite{Dinh2012}. This method principally replaces BMIs with a difference of convex functions, which can subsequently be transformed into LMIs using the first-order Taylor approximation and the Schur complement. 
If the optimal value of the optimization problem~\eqref{equ:LINF} is denoted by $f^*$, then the approximating convex problem has optimal value $\bar{f}^*$ such that $f^* \leq \bar{f}^*$. 
The next theorem presents the result from applying the SCA to \eqref{equ:LINF} and hence yielding a sequence of convex routines to solve.
\begin{theorem}\label{th2}
The convex approximation of problem~\eqref{equ:LINF} around the point $(\tilde{\alpha},\tilde{\mu}_0,\tilde{\mu}_1,\tilde{\m S})$ can be written as an SDP with optimization variables $\m S, \m Z$ and positive scalars $\mu_0, \mu_1, \mu_2, \alpha, \hslash$
\begin{subequations}~\label{equ:LINF-Niko} 
	\begin{align}
	\min & \;\;\; \hslash \\
	\st & \;\;\; \mathcal{\m C}_t(\m S,\m Z, \alpha, \mu_0, \mu_1, \mu_2, \hslash;\tilde{\m S}, \tilde{\alpha},\tilde{\mu}_0,\tilde{\mu}_1 ) \preceq 0
	\end{align}
\end{subequations}
where $\mathcal{\m C}_t(\cdot)$  is a block diagonal concatenation of the LMI constraints defined in Appendix~\ref{proofth2}.
	\end{theorem}
Proof of Theorem~\ref{th2} is in Appendix~\ref{proofth2} alongside the closed form representation of $\mathcal{\m C}_t(\cdot)$. 
As stated earlier, this method relies on the SCA of the nonconvex constraints around a linearization point.  Let $k$ be the index of a problem that is solved in every iteration; and let $\alpha_k,\,\mu_{0_k},\,\mu_{1_k},\,\mu_{2_k},\,\hslash_k,\,\m S_k,\,\m Z_k$ be the corresponding solution. An additional term $J_k$ in the objective function of~\eqref{equ:LINF-Niko} is added to improve convergence. This term can be written as $$ J_k = \Vert\alpha-\tilde{\alpha} \Vert _2^2+\Vert\mu_{0}-\tilde{\mu}_0 \Vert_2^2+\Vert\mu_{1}-\tilde{\mu}_1 \Vert_2^2+\Vert\m S-\tilde{\m S} \Vert_F^2$$ 
where $\tilde{\alpha} = \alpha_{k-1}$, $ \tilde{\mu}_0 = \mu_{0_{k-1}}$, $\tilde{\mu}_1 = \mu_{1_{k-1}} $, $\tilde{\m S} = \m S_{k-1} $. The $k$-th SCA of~\eqref{equ:LINF} can be written as
	\begin{eqnarray}\label{equ:LINF-Niko-Seq}
	\bar{f}^*_k = \minimize && \hslash_k + \gamma J_k\\
\subjectto	& &{\small\mathcal{\m C}_t(\m S,\m Z, \alpha, \mu_{0_{}}, \mu_{1_{}}, \mu_{2_{}}, \hslash;\tilde{\m S}, \tilde{\alpha},\tilde{\mu}_0,\tilde{\mu}_1 ) \preceq 0} \notag 
	\end{eqnarray}
where $\alpha,\,\mu_{0},\,\mu_{1},\,\mu_{2},\,\hslash,\,\m S$, and $\m Z$ are the new optimization variables; $\gamma>0$ is a regularization weight; and $\bar{f}^*_k$ is the optimal value of~\eqref{equ:LINF-Niko-Seq} at the $k$-th SCA iteration. Algorithm~\ref{algorithm:SCA} provides the steps to solve~\eqref{equ:LINF-Niko-Seq} sequentially until a maximum number of iterations ($\mathrm{MaxIter}$) or a stopping criterion defined by a tolerance ($\mathrm{tol}$) is achieved.

Based on the general framework of~\cite{Dinh2012}, Algorithm~\ref{algorithm:SCA} enjoys several convergence properties. In particular, the sequence $\{\bar{f}^*_k\}$ is monotonically decreasing; and by construction, it is an upper bound to $f^*$ [cf.~\eqref{equ:LINF}]. Furthermore, under mild regularity conditions listed in~\cite{Dinh2012}, every accumulation point of the sequence of solutions $\{\alpha_k,\,\mu_{0_k},\,\mu_{1_k},\,\mu_{2_k},\,\hslash_k,\,\m S_k,\,\m Z_k\}$ to~\eqref{equ:LINF-Niko-Seq} is a KKT point of~\eqref{equ:LINF}. 

\begin{rem} \label{SCAinit1}
	Since the SCA is an inner approximation of the nonconvex problem, it needs to start from a strictly feasible point. 
To obtain this point, $\alpha$ and $\mu_{1}$ can always be set to a desired predefined values and then solve problem \eqref{equ:LINF} as an SDP with LMI constraints.  
\end{rem}

After the implementation of Algorithm~\ref{algorithm:SCA}, the state-feedback control is computed as $\m u = \m K^* \m x(t)$ where $\m K^* = \m Z^*(\m S^*)^{-1}$. Per Theorem~\ref{th1}, this gain guarantees that $$ \Vert  \m z(t) \Vert_2=  \Vert \m C\m x(t) + \m D \m u(t) \Vert_2 \leq \mu^* \rho$$
for all $t>t_0$, where $\mu^* = \sqrt{\mu_0^*\mu_1^*+\mu_2^*}$ and $\rho = \Vert \m w \Vert_{\mc{L}_\infty}$ which can be considered as the worst-case disturbance. Section~\ref{sec:numtests} explores whether these performance guarantees hold under various conditions. 
\begin{algorithm}[!h] 
	\caption{Solving the SCA of~\eqref{equ:LINF}.}
	\label{algorithm:SCA}
	\begin{algorithmic}
		\STATE \textbf{initialize:} $k=1$, $\alpha_{k-1}$, $\mu_{0_{k-1}}$, $\mu_{1_{k-1}}$, and $\m S_{k-1}$
		\WHILE {$k < \mathrm{MaxIter} $}
		\STATE Solve~\eqref{equ:LINF-Niko-Seq} 
		\IF{ $|\bar{f}^*_k-\bar{f}^*_{k-1}| < \mathrm{tol}$ }
		\STATE \textbf{break}
		\ELSE 
		\STATE $k\leftarrow k+1$
		\ENDIF		\ENDWHILE
		\STATE {\small\{$\m S^\star, \m Z^\star \} \leftarrow \{\m S_k,\m Z_k\}$,} $\m K^\star \leftarrow \m Z^\star(\m S^\star)^{-1}$
	\end{algorithmic}
\end{algorithm}

\section{Decentralized $\mathcal{L}_{\infty}$ Control Formulation}\label{sec:decentralized}
The formulation presented in the previous section assumes a centralized control law, that is, matrix $\m K$ is dense, which is practical in microgrids or in areas where utilities or system operators have full access to the network's states. This assumption is reasonable in future power networks with increased installations of PMUs and dynamic state estimation methods. 
In this section, we present a decentralized controller that ensures that each local controller only uses locally acquired measurements. Specifically, the two local control signals for each generator only require the knowledge of generator's states. The design can also be extended to multi-area power networks with each area having the measurements from all the buses in that area. 

Following a similar derivation of Theorem~\ref{th1} for the centralized $\mc{L}_{\infty}$ controller, the robust $\mc{L}_{\infty}$ decentralized control problem can be derived considering that $\m K$ is an optimization variable instead of computing it from the resulting matrices $\m Z$ and $\m S$. This is then followed by imposing strict structure on $\m K$ that defines the decentralized control architecture---be it purely decentralized or distributed.
In particular, in the proof of Theorem~\ref{th1} we do not apply the congruence transformation and the change of variables before~\eqref{LMI1}, but keep $\m K$ as an optimization variable. The resulting formulation is 
\begin{subequations}~\label{equ:LINFDec}
	\begin{align}
\min& \;\;\;\mu_0 \mu_1 + \mu_2 \\
		\st & 	\begin{bmatrix}
			(\m A^{\top}+\m K^{\top}\m B_u^{\top})\m P\\+\m P(\m A+\m B_u\m K)+\alpha \m P & \m P\m B_w \\
		\star & -\alpha \mu_0 \m I
		\end{bmatrix} \preceq 0 ~\label{equ:LINFDec-BMI1}\\
		& \begin{bmatrix}
			-\mu_1\m P& \m O &\m C+\m D\m K \\ \star & -\mu_2\m I & \m O \\ \star& \star & -\m I
		\end{bmatrix}
		\preceq 0  ~\label{equ:LINFDec-BMI2}\\
		&\bmat{-\mu_0\rho^2 &\m P\m x_0 \\ \star & -\m P} \preceq 0,\;\bmat{-\frac{u_{\max}^2}{\rho^2}\m P & \mu_0\m K\\ \star& -\mu_0 \m I} \preceq 0~\label{equ:LINFDec-BMI3}\\
		&\;\m K \in \mathcal{K},
	\end{align}
\end{subequations}
where the optimization variables are $\m P$, $\m K$ and the positive scalars. The constraint $\m K \in \mathcal{K}$ defines the convex set that describes the decentralized control architecture. For example, if purely decentralized controllers are sought, then $K_{ij}=0$  can be included in $\mathcal{K}$ for all $(i,j)$ except for the ones representing feedback of local measurements to local controler inputs.  Remark~\ref{rem:Dec} includes a discussion on this constraint. 

Similar to the SCA and derivations in the previous section, problem~\eqref{equ:LINFDec} can be solved using a specific successive convex approximation, which is detailed in the next theorem. We do not consider the input bound constraints as a part of the SCA for the decentralized $\mc{L}_{\infty}$ due to the lack of space.
\begin{theorem}\label{th:dec}
	The convex approximation of problem~\eqref{equ:LINFDec} around the point $(\tilde{\alpha},\tilde{\mu}_0,\tilde{\mu}_1,\tilde{\m P},\tilde{\m K})$ can be written as the following SDP with optimization variables $\m P, \m K$ and positive scalars $\mu_0, \mu_1, \mu_2, \alpha, \hslash$:
	\begin{subequations}\label{equ:LINFDec-Niko2}
		\begin{align}
			&\min \;\;\;\hslash \\
			\st&\begin{bmatrix}
				\m \Xi & \star & \star & \star & \star \\ 
				\m B_w^{\top}\m P & \frac{1}{4}F_{l_2}(\cdot) & \star & \star & \star \\
				\frac{1}{2}(\alpha\m I+\m P) & \m O & -\m I & \star & \star \\
				\m O & \frac{1}{2}(\alpha-\mu_0)\m I & \m O & -\m I & \star \\
				\frac{1}{\sqrt{2}}(\m P + \m B_u \m K) & \m O & \m O & \m O & -\m I
			\end{bmatrix} \preceq 0 ~\label{equ:LINFDec-LMI12}\\
			&\begin{bmatrix}
				\frac{1}{4}H_{l}(\cdot) & \star & \star & \star\\ \m O & -\mu_2\m I & \star & \star\\ \m C +\m D \m K& \m O & -\m I & \star \\ \frac{1}{2}(\mu_1\m I - \m P) & \m O & \m O & -\m I
			\end{bmatrix}
			\preceq 0  ~\label{equ:LINFDec-LMI22}\\
			&\begin{bmatrix}
				\frac{1}{4}H_{l}(\cdot)+\mu_2-\hslash & \star \\ \frac{1}{2}(\mu_0+\mu_1) & -1
			\end{bmatrix} \preceq 0,\; \m K \in \mathcal{K}, ~\label{equ:LINFDec-calK2}
		\end{align}
	\end{subequations}
where $\m \Xi = 	\m A^{\top}\m P+\m P\m A +\frac{1}{2}G_{l}(\cdot) +\frac{1}{4} F_{l_1}(\cdot)$ and $F_{l_1},F_{l_2},G_{l}, H_{l}$ are all linear matrix-valued functions of the optimization variables given in Appendix~\ref{app:DLINF}.
\end{theorem}
The proof of Theorem~\ref{th:dec} and the closed form expressions of the linear matrix-valued functions are all presented in Appendix~\ref{app:DLINF}.
An SCA algorithm akin to Algorithm~\ref{algorithm:SCA} can also be implemented to obtain $\m K^*, \mu_0^*, \mu_1^*, $ and $\mu_2^*$ yielding the desired $\mc{L}_{\infty}$ performance level $\mu^*=\sqrt{\mu_0^*\mu_1^*+\mu_2^*}$ for the decentralized control architecture. Note that the nonconvexity in \eqref{equ:LINFDec} is different from that of \eqref{equ:LINF} as the bilinearities only appear as multiplications between scalar variables or between scalar and matrix variables. For the decentralized $\mc{L}_{\infty}$ formulation, one of the bilinearities appears as a multiplication of two matrix variables $\m P$ and $\m K$, where $\m K \in \mathcal{K}$. This type of bilinearity makes it challenging to obtain a strictly feasible point of problem \eqref{equ:LINFDec}, which is needed as an initialization. To that end, we develop an algorithm to initialize the SCA for \eqref{equ:LINFDec} based on the methods in~\cite[Section V]{Dinh2012}. 
\begin{rem}~\label{rem:Dec}
The convex constraint $\m K \in \mc{K}$ can be arbitrarily chosen by the system operator, as it depends on the logistics of the controller. If a purely decentralized controller is desired, then the $i$-th controller only measures its generator states.  
\end{rem}

\vspace{-0.4cm}

\color{black}
\section{Numerical Experiments}~\label{sec:numtests}
In this section, numerical simulations are presented to  investigate the application of the aforementioned algorithms in stabilizing several standard  IEEE test networks under load and renewable disturbances. The SDPs are modeled via YALMIP~\cite{lofberg2004yalmip} and solved by MOSEK~\cite{mosek2015mosek}.
The operating point of the  power network $(\m x^0, \m u^0, \m a^0)$ is obtained given  $\m w^0$ using optimal power flow. The  linearized state-space parameters are then computed.  Next, the $\mathcal{L}_{\infty}$ feedback gain $\m K$ is calculated via Algorithm~\ref{algorithm:SCA} and Theorem~\ref{th2}. The feedback controller $\m u(t)=\m K\m x(t)$ is applied to the nonlinear power network given in~\eqref{SynGen},~\eqref{SynGenPF},~\eqref{GPF}, and~\eqref{GPFe}. The nonlinear DAEs are simulated via MATLAB's ODE suite. The objective of this section is two-fold: Comparing the performance of the $\mc{L}_{\infty}$ controller with other control methods in the literature of power networks under various conditions, and investigating whether the performance bounds from Theorem~\ref{th1} hold. 
\subsection{Power System Parameters and Setup}
The 9-bus system, 39-bus New England system, and a 57-bus  system are  selected to conduct the numerical simulations. The steady-state data required to construct the power flow equations in \eqref{GPF} are obtained from MATPOWER~\cite{zimmerman2011matpower}.    
Synchronous machine constants for characterizing generator dynamics  based on the fourth-order model in \eqref{SynGen} are obtained  from Power System Toolbox case files \texttt{d3m9bm.m}, \texttt{datane.m}  for the 39-bus network~\cite{sauerpower}.   For the 57-bus network, as well as the governor model of~\eqref{SynGen4} for all networks, typical parameter values of $M_i=0.2~\mr{pu} \times \mr{sec}^2$, $D_i=0~\mr{pu} \times \mr{sec}$, $\tau_{d_i}=5~\mr{sec}$, $x_{d_i}=0.7~\mr{pu}$, $x_{q_i}=0.5~\mr{pu}$, $x'_{d_i}=0.07~\mr{pu}$, $\tau_{c_i}=0.2~\mr{sec}$, and $R_i=0.02~ \frac{\mr{Hz}}{\mr{pu}}$ have been selected based on ranges of values provided in PST.  For later reference, the power base is $100$ MVA. The total initial load, that is $\sum_{n \in \mc{N}} p_{l_n}^0$, is $3.15+j1.15~\mr{pu}$, $62.54+j13.87~\mr{pu}$, and $12.51+j3.36~\mr{pu}$ for the 9-,  39-, and 57-bus networks, respectively.
When wind farms are added to the standard test cases, we set $\mc{R}=\mc{G} \cup \mc{L}$ so that $\mc{R}=\mc{N}$. Wind injection is modeled as negative loads, effectively injecting power into the network albeit reducing system inertia.

\subsection{Robust $\mc{L}_{\infty}$ vs. Automatic Generation Control}
Here, we evaluate  the performance of the proposed $\mc{L}_{\infty}$ controller on the 39-bus network for the centralized architecture of feedback control and compare it with automatic generation control (AGC) under two settings: \textit{(a)} a step disturbance in load but without wind generation and \textit{(b)}  a step disturbance in load with wind generation.  
The performance index is selected by setting $\mC_{\m \delta, \m \omega, \m e}=0.2 \m I$, $\mC_{\m m} =0.1\mI$, $\mD_{\m r}=0.1 \mI$, and $\mD_{\m f}= 0.2 \mI$.  Appendix \ref{Appendix:AGC} includes the  AGC implementation.

\begin{figure}[t]
	\vspace{-0.5cm}
\color{black}	\subfloat[]{\includegraphics[scale=0.22]{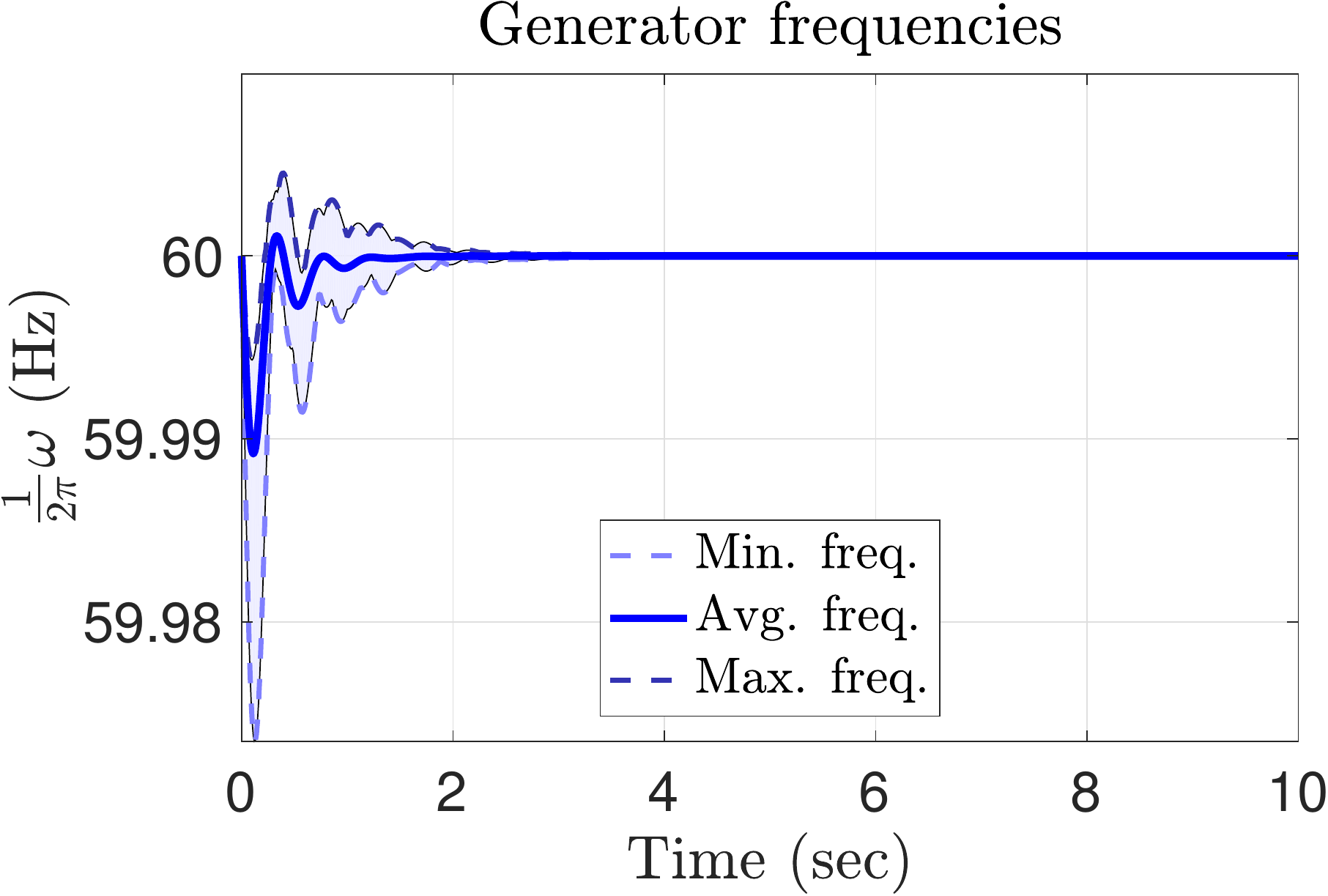}\label{Fig:CentFreqNoWind}} 
	\subfloat[]{\includegraphics[scale=0.22]{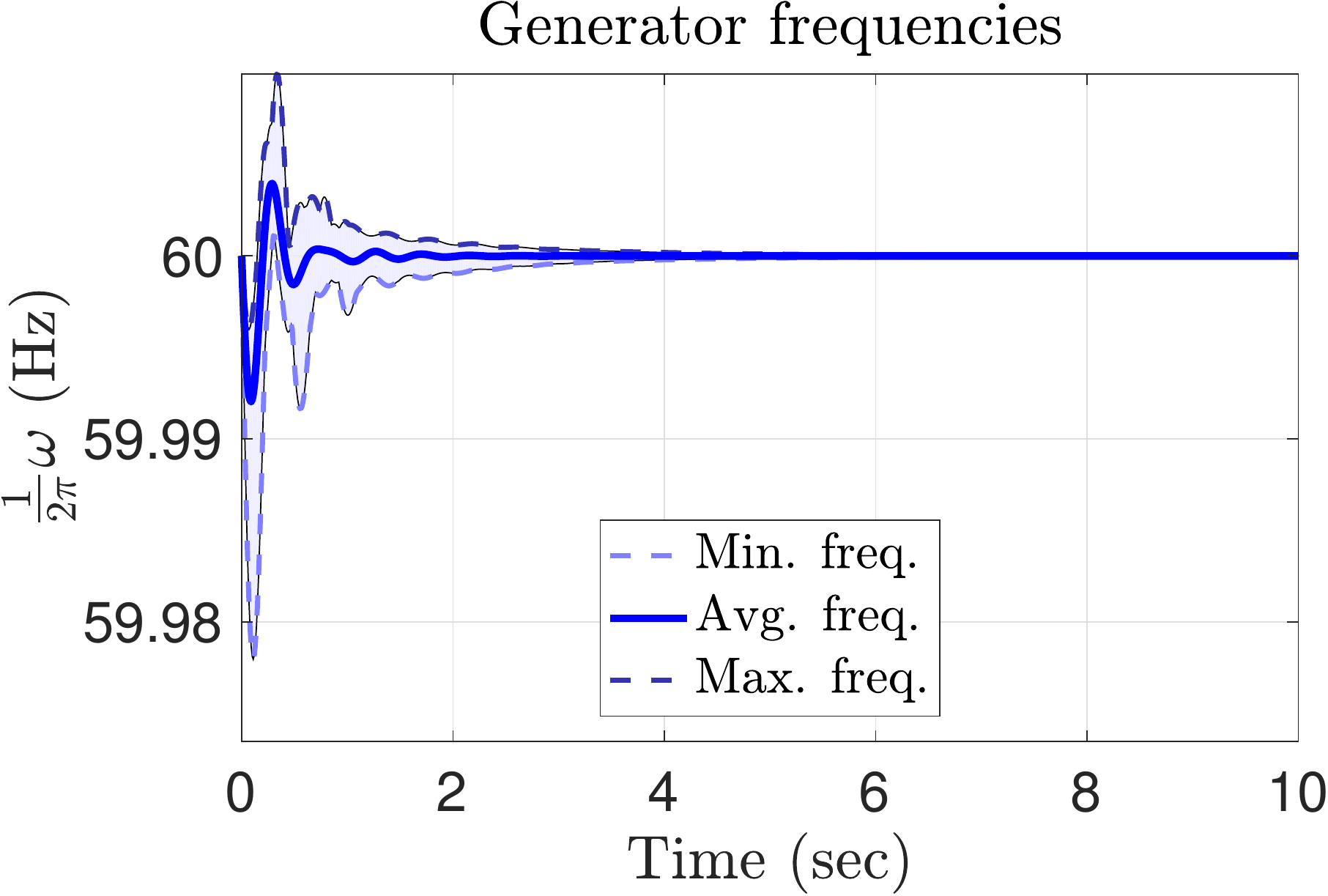}\label{Fig:AgcFreqNoWind}} \\
	\caption{Frequency performance under a load step disturbance in a power network without wind generation: (a) $\mc{L}_{\infty}$ controller and  (b) AGC.   Both controllers manage to control the frequency around the nominal value.  \label{Fig:CentAgcFreqNoWind}}
\end{figure}

\textcolor{black}{The system initially operates with total load of $(p_{l_n}^0, q_{l_n}^0)$ and zero wind capacity. For $t>0$, a sudden step-change occurs in the load. In particular, the value of $p_{l_n}(t)$ varies as follows:
\begin{IEEEeqnarray}{rCl}
	p_{l_n}(t)&=& p_{l_n}^0+ \Delta p_{l_n}, n \in \mc{N} \IEEEyesnumber \label{Eq:Steppl} 
\end{IEEEeqnarray}
where we set $\Delta p_{l_n}=0.03 p_{l_n}^0$ as a step change.  Step disturbances in power system frequency control studies is a common practice~\cite{ZhaoTopcuLiLow2014}. }
For a time span of $t \in [0, 10]$ seconds, the disturbance in~\eqref{Eq:Steppl} causes the DAEs to depart from the initial equilibrium.  This disturbance corresponds to a  signal $\m w(t)$ with $\| \m w\|_{\mc{L}_\infty}=   0.1674   ~\mr{pu}$.  The centralized feedback law previously computed is then applied to the nonlinear, perturbed power network. The behavior of the nonlinear dynamical system is then analyzed.   The frequency response in this setup is given in Fig.~\ref{Fig:CentAgcFreqNoWind} for both the robust $\mc{L}_{\infty}$ controller and AGC; the figure shows the ranges of frequencies for all buses. 

\begin{figure}[t]
\color{black}
\centering 	
{\includegraphics[scale=0.3]{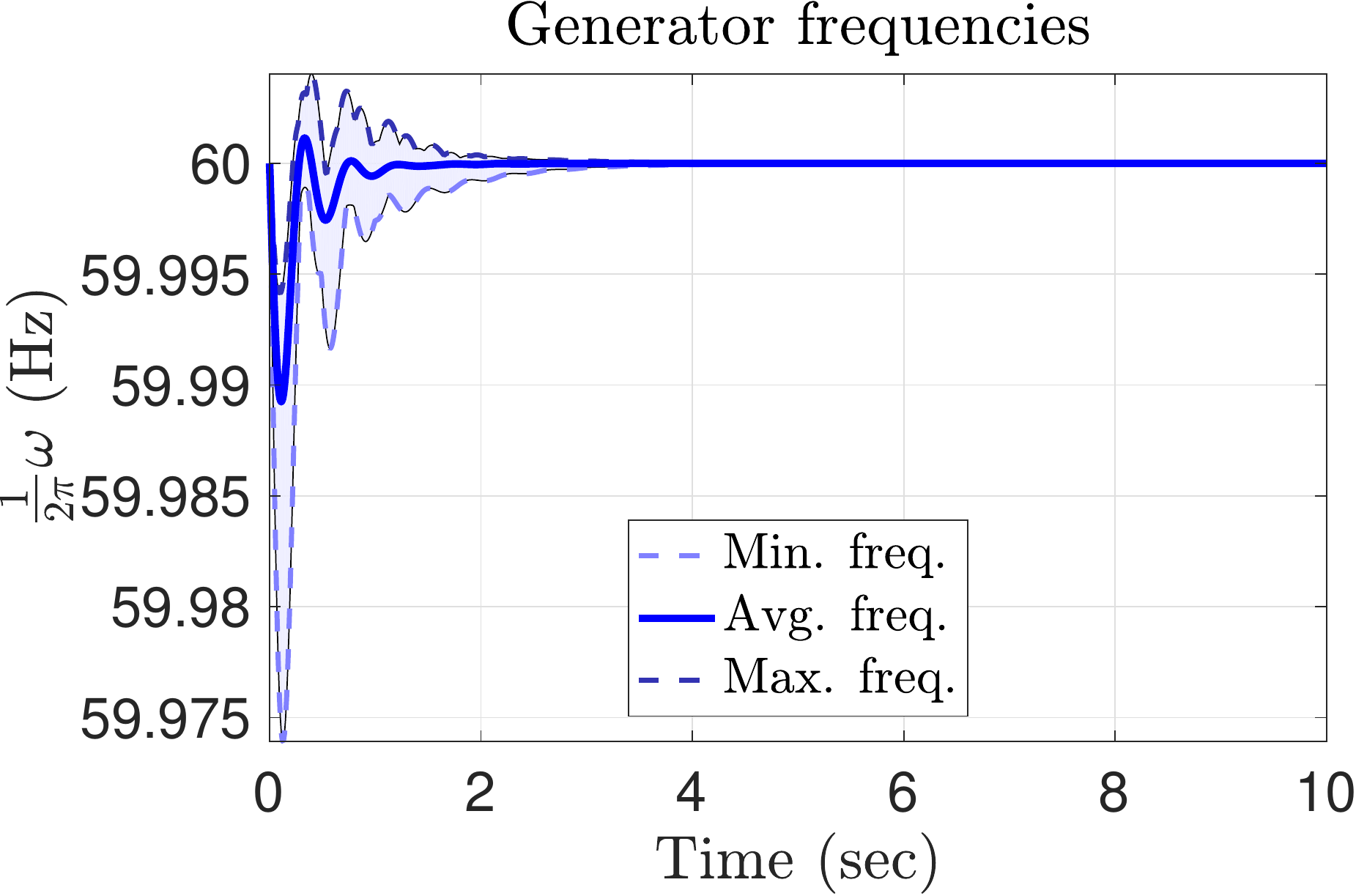}\label{Fig:CentFreqWind}} 
	\caption{Frequency performance under a load step disturbance in a power network with wind generation for the $\mc{L}_{\infty}$ controller.  It is evident that the $\mc{L}_{\infty}$ controller manages to control the frequency around the nominal value. AGC fails to produce stable frequency oscillations under heavy wind disturbance. \label{Fig:CentAgcFreqWind}}
			\vspace{-0.5cm}
\end{figure}

 In our implementation of the AGC, the governor is controlled according to  \eqref{Eq:prefParticipation} in Appendix~\ref{Appendix:AGC} while the exciter uses the optimal  $\mc{L}_{\infty}$ feedback signal to aid voltage control.
In an additional setup, wind farms are made responsible for  $0.2p_{l_n}^0$ of generation.  The same disturbance~\eqref{Eq:Steppl} is applied again and dynamical results are recorded. The frequency response in this case is depicted in Fig.~\ref{Fig:CentAgcFreqWind}. 
Maximum frequency deviation for both setups are also printed in Table~\ref{Table:CentAGCSummary} for $\mc{L}_{\infty}$ and AGC control methods.   It is observed that when the low-inertia wind generators are included, AGC fails in frequency control under a step-change in load and significant wind generation---and the corresponding frequency plots diverge---whereas  $\mc{L}_{\infty}$ is successful in stabilizing the grid's frequency.

\begin{table}[t]
		\vspace{-0.5cm}
\color{black}
	\centering
	\renewcommand{\arraystretch}{1}
	\caption{\small Maximum frequency deviation comparison between $\mc{L}_{\infty}$ and AGC controllers for the 39-bus network.}
	\label{Table:CentAGCSummary}
	\begin{tabular}{|c|c|c|c|}
		\hline
		Controller       &Freq. Dev. (No wind)  & Freq. Dev (Wind)  \\
		\hline\hline
		$\mc{L}_{\infty}$                        &     0.0265    (Hz)     &    0.0261   (Hz)      \\ 
		\hline       
		AGC                                     &  0.0221    (Hz)    & 0.4764    (Hz)      \\
		\hline      	
	\end{tabular}
\end{table}

\begin{table}[t]
\color{black}
	\centering
	\renewcommand{\arraystretch}{1}
	\caption{\small Performance evaluation of  $\mc{L}_{\infty}$, and $\mc{H}_{\infty}$ controllers for the 39-bus network.}
	\label{Table:CentHinfSummary}
	\begin{tabular}{|c|c|c|}
		\hline
		Controller       &Max.  Freq. Dev.   & Max.  Volt. Dev.  \\
		\hline\hline
		$\mc{L}_{\infty}$                        &    0.0415  (Hz)   &  0.0472  (pu)      \\ 
		\hline       
		$\mc{H}_{\infty}$           &  0.0441   (Hz)        & 0.1147   (pu)      \\
		\hline      	
	\end{tabular}
\end{table}
\begin{figure}[t]
\color{black}
	\centering
	\includegraphics[scale=0.35]{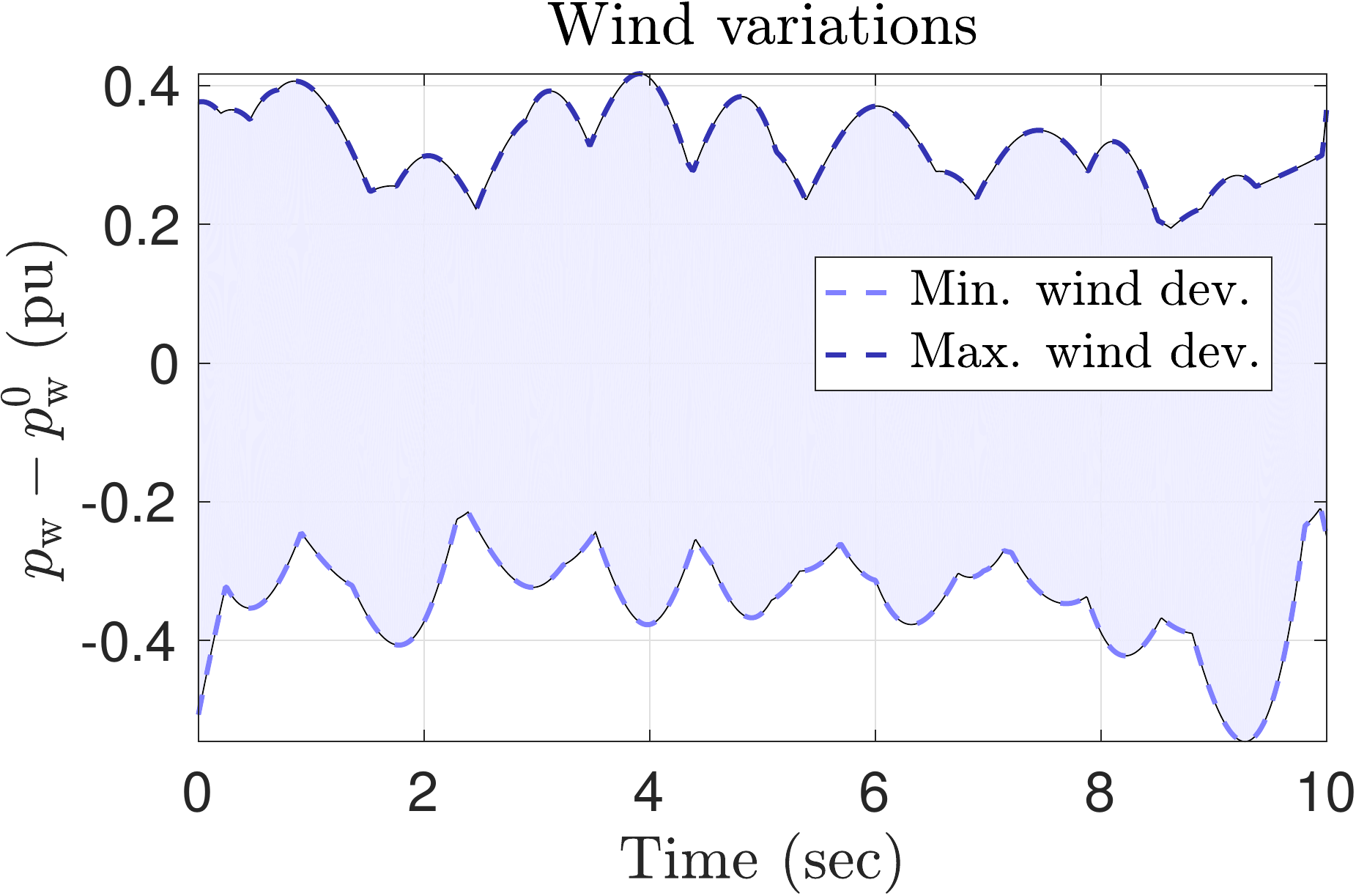}
	\caption{ Range of wind variations around the predicted value of $p_{\mathrm{w}}^0$. \label{Fig:SampleWind}}
\end{figure}
\begin{figure}[t]
\color{black}
	\subfloat[]{\includegraphics[scale=0.23]{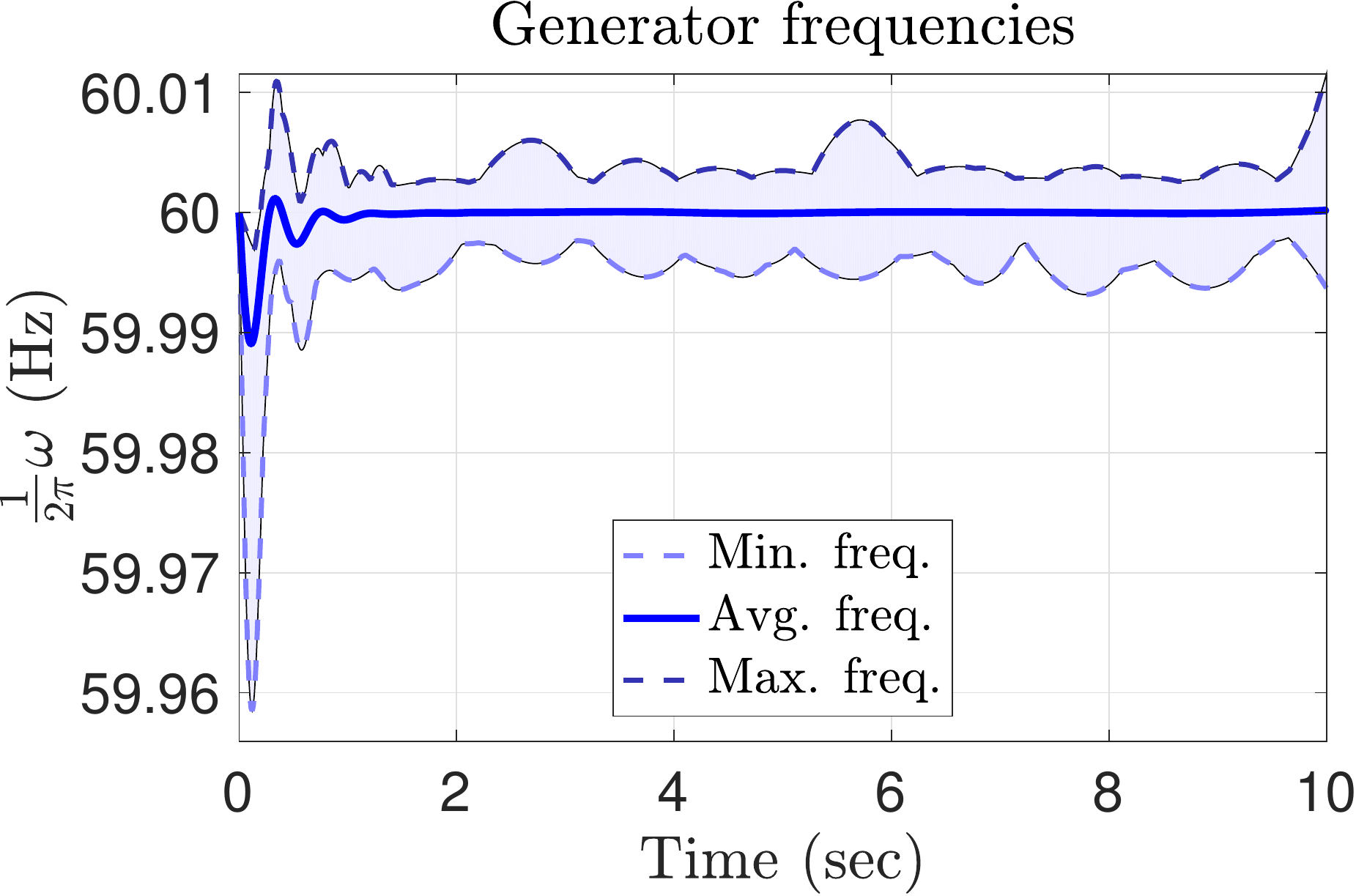}\label{Fig:CentFreq}} 
	\subfloat[]{\includegraphics[scale=0.23]{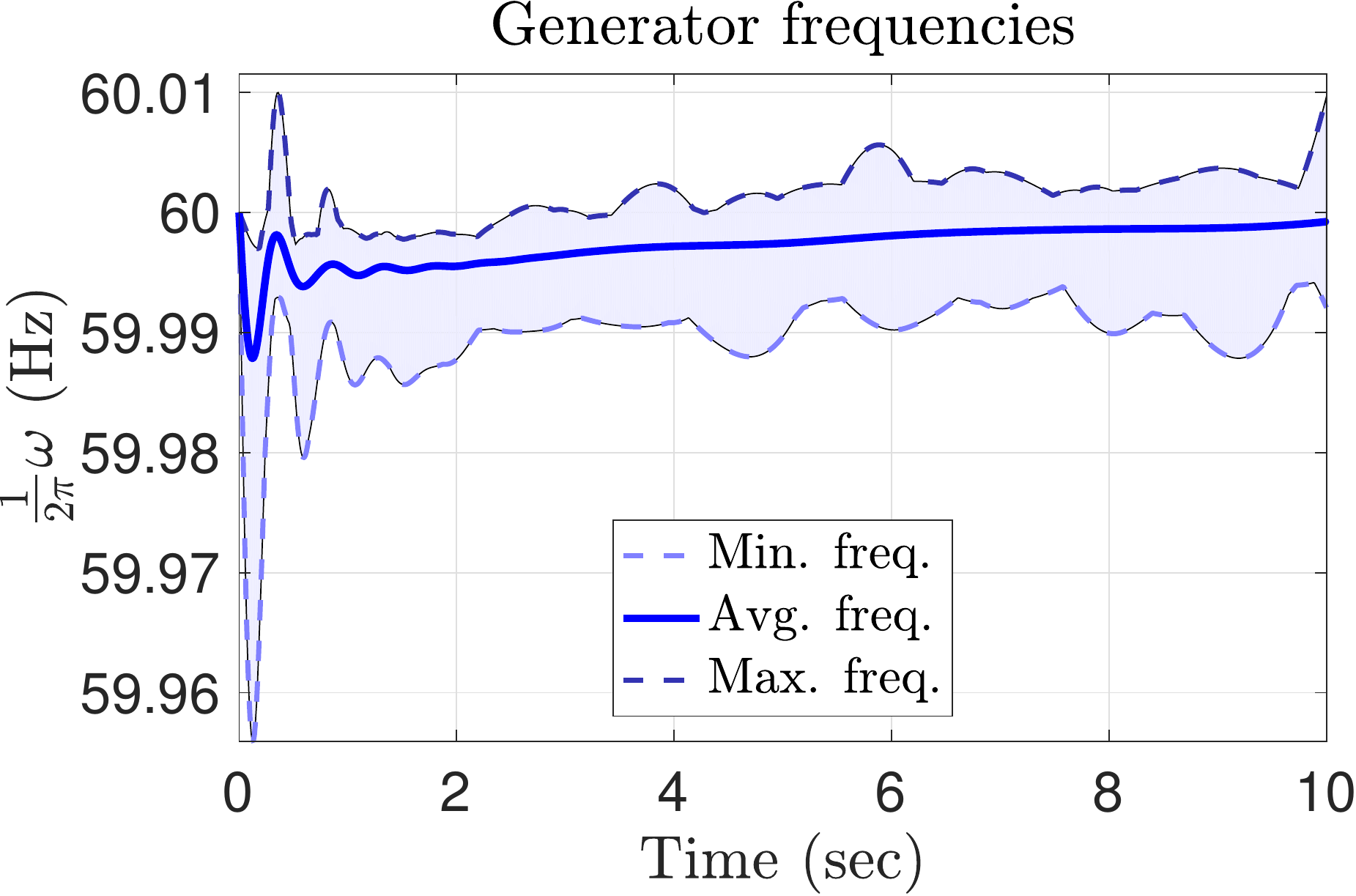}\label{Fig:HinfWind}} \\
	\subfloat[]{\includegraphics[scale=0.23]{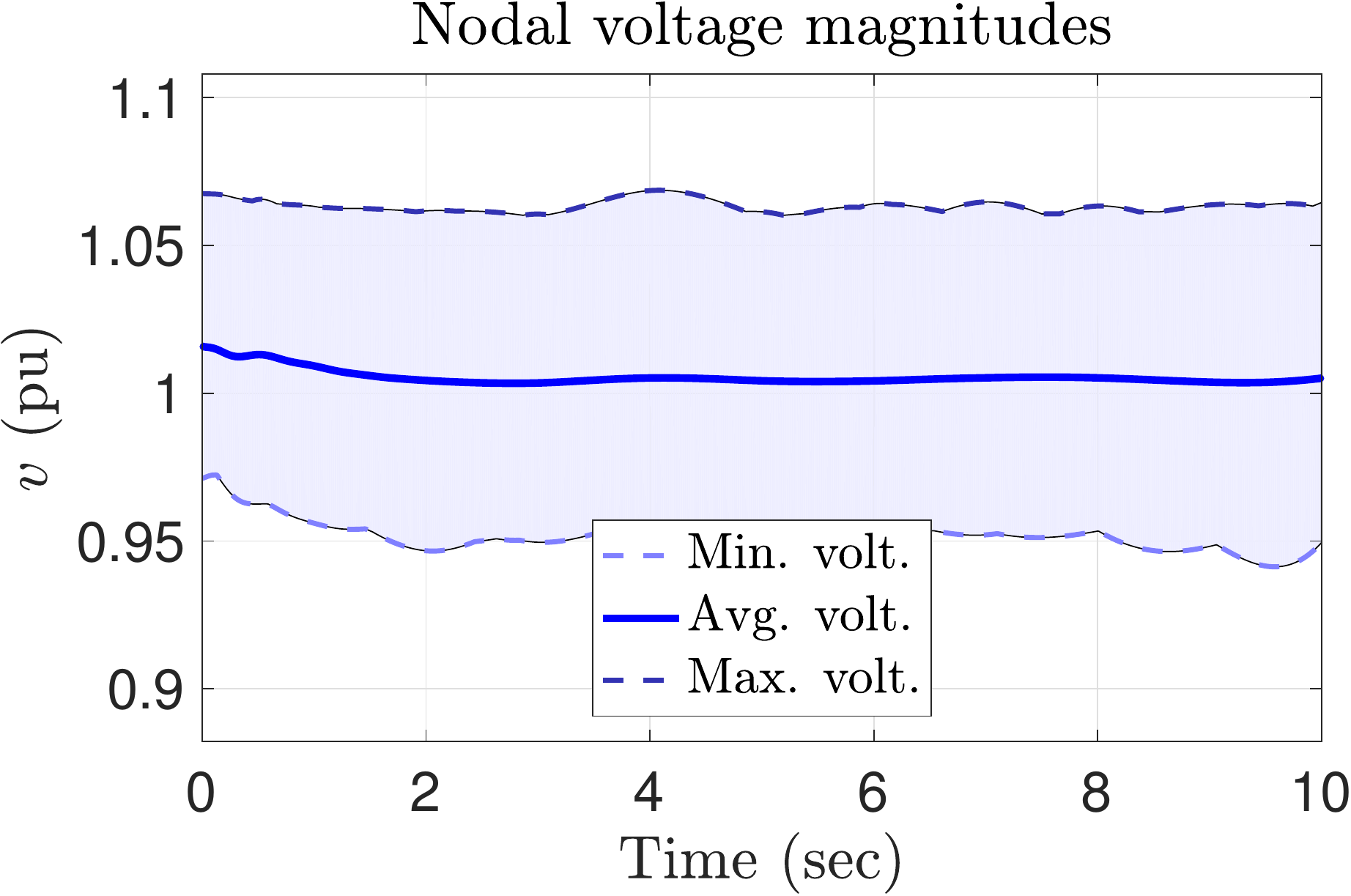}\label{Fig:CentVMags}} 
	\subfloat[]{\includegraphics[scale=0.23]{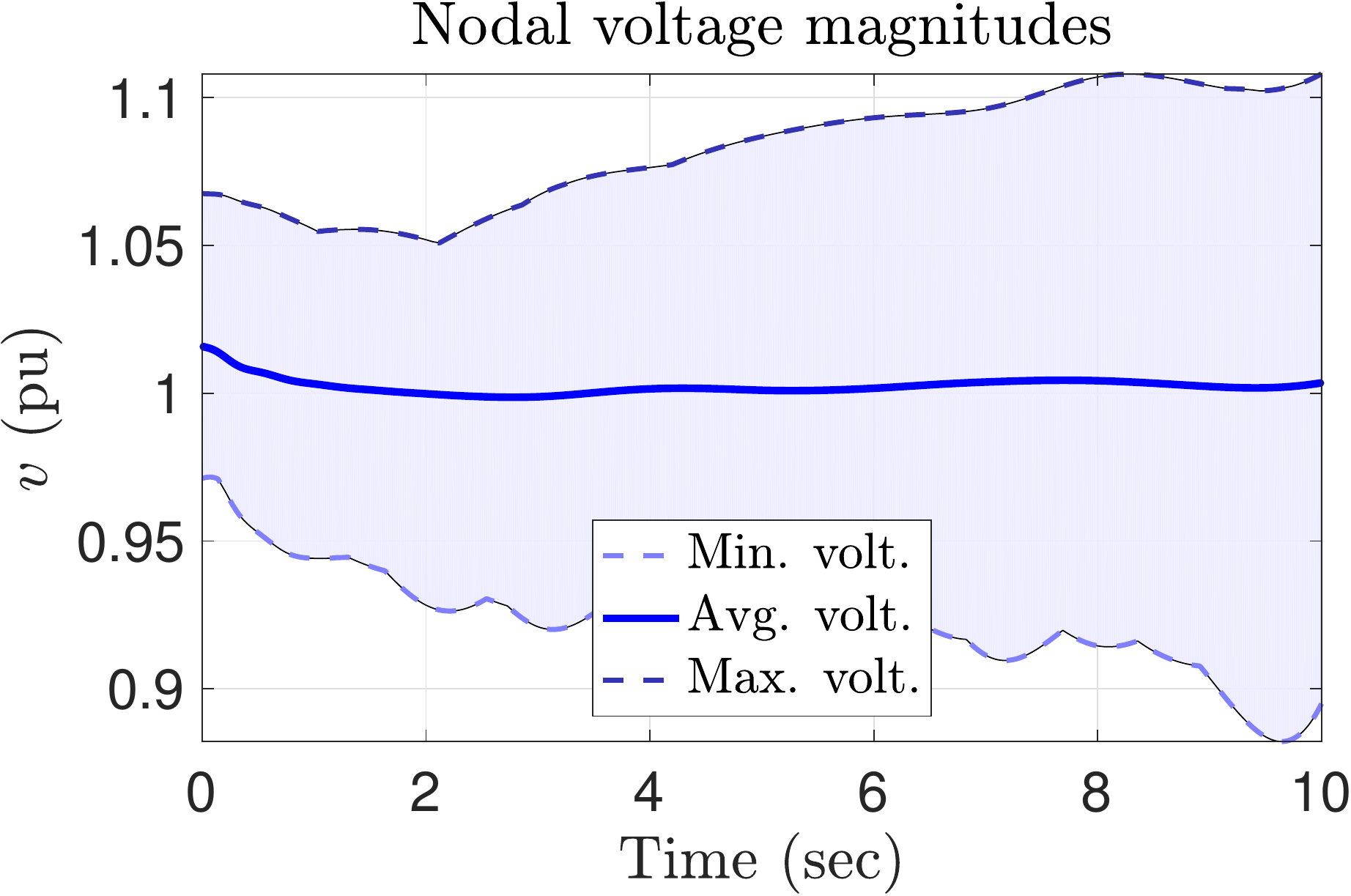}\label{Fig:HinfVMags}} \\
	\caption{Range and averages of frequency and voltage plots for 50 wind realizations and across all buses. Control performance under a load step disturbance and large wind variation: for $\mc{L}_{\infty}$ controller (left) and  $\mc{H}_{\infty}$ controller (right).   \label{Fig:CentHinf}}
			\vspace{-0.5cm}
\end{figure}

\subsection{\textcolor{black}{Robust $\mc{L}_{\infty}$ vs. $\mc{H}_{\infty}$ control}}
\textcolor{black}{In this section, we  compare the $\mc{L}_{\infty}$ with the robust  $\mc{H}_{\infty}$ controller implemented via the  LMI formulation given in~\cite{gahinet1994linear} and later used in power networks. This implementation is shown in Appendix~\ref{Appendix:AGC}.}
Similar to the second setup in the previous section,  wind farms are made responsible for  $0.2p_{l_n}^0$ of generation. \textcolor{black}{In this setup,  the step disturbance in load model~\eqref{Eq:Steppl} is further varied to incorporate random load and wind variations together, as follows:
\begin{IEEEeqnarray}{rCl}
	\IEEEyesnumber \label{EqGroup:NoiseVariations}
	p_{l_n}(t)&=& p_{l_n}^0+ \Delta p_{l_n}+z_{l_n}(t), n \in \mc{N} \IEEEyessubnumber \label{Eq:TimeVaryingpl}  \\
	p_{r_n}(t)&=& p_{r_n}^0+z_{r_n}(t), n \in \mc{R} \IEEEyessubnumber \label{Eq:TimeVaryingpr}
\end{IEEEeqnarray}
where  quantities $z_{l_n}(t)$ and $z_{r_n}(t)$ are Gaussian noise with zero mean and variance of $0.33 \Delta p_{l_n}$ and large wind variance of $0.05 p_{r_n}^0$ (see also~\cite[Section IV.D]{LeXie2011} for  similar levels of wind variations).}  The simulations  are repeated for 50 such random realizations.   The range of wind variation is shown in Fig.~\ref{Fig:SampleWind}.  Results from trajectories are recorded in Table~\ref{Table:CentHinfSummary}. We find that both methods have a satisfactory frequency control performance while the $\mc{L}_{\infty}$ control method outperforms the $\mc{H}_{\infty}$ in voltage control performance.  Notice that frequency and voltage deviations are maximum deviations  measured with respect to the initial point $\m x^0$ and across all 50 wind realizations.   Ranges and averages of frequency/voltage plots for all realizations for all buses are  provided in Fig.~\ref{Fig:CentHinf}. 

\subsection{Decentralized and Input-Constrained  $\mc{L}_{\infty}$ Control}
In this section, we evaluate the performance of the decentralized and input-constrained designs in comparison to the centralized controller in stabilizing the 9-, 39-, and 57-bus networks (in the previous section, input constraints are not imposed).  For the input-constrained problem, we select $u_{\max}=5$ (pu).  For the decentralized controller, $\mc{K}$ is selected so that only local measurements are used to compute the local input feedback, effectively enforcing a block-diagonal feedback structure on the control gain $\m K$.

\begin{figure}[t]
	\color{black}
	\centering 	
	{\includegraphics[scale=0.4]{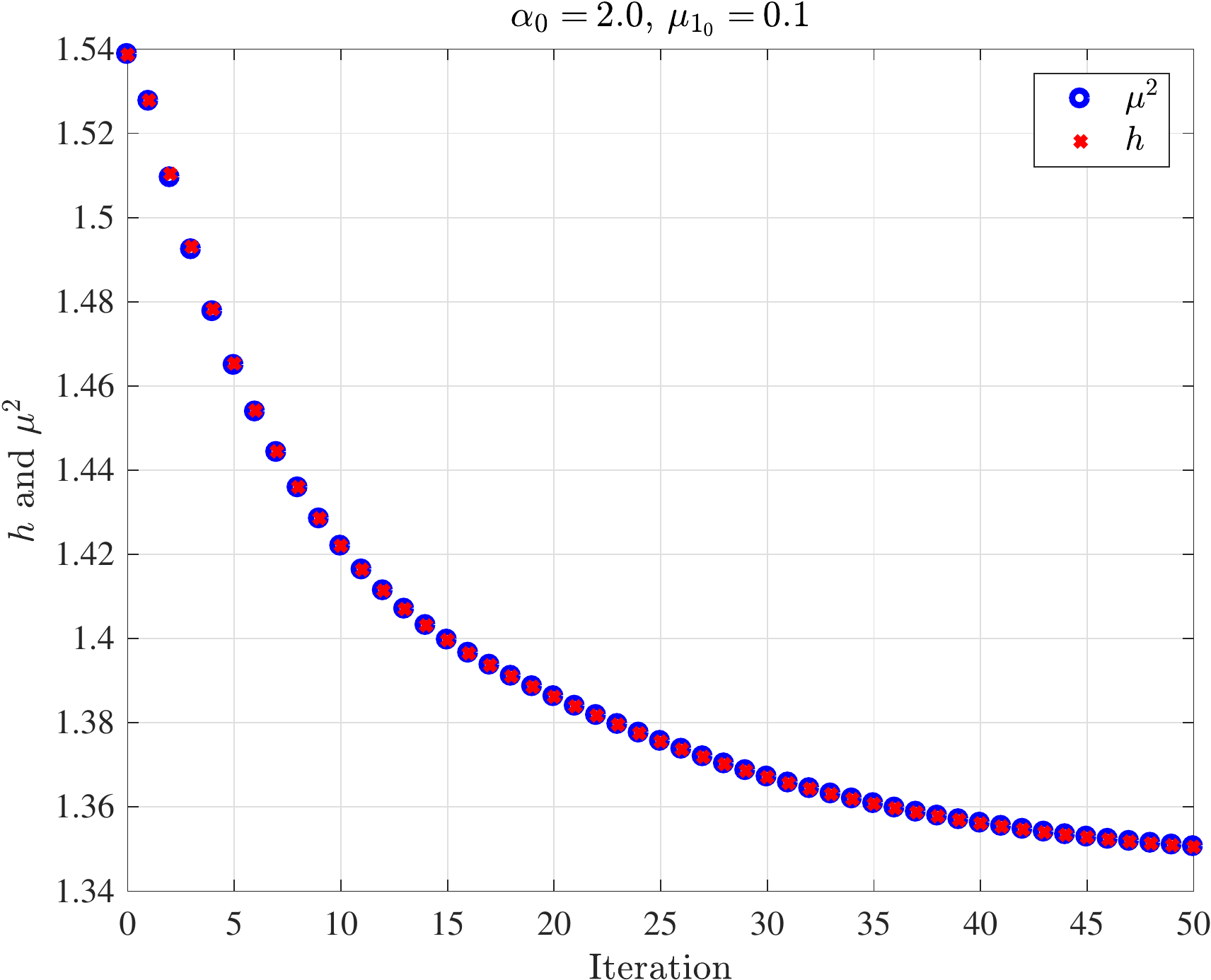}} 
	\caption{\textcolor{black}{Iterations of the SCA algorithm to compute a  centralized stabilizing $\m K$ for the 57-bus network.}}
	\label{Fig:Hmu}
\end{figure}

The disturbances applied are of the form~\eqref{EqGroup:NoiseVariations} with $\mc{L}_{\infty}$-norms of  $\| \m w \|_{\mc{L}_\infty}=    0.0861$ (pu), $\| \m w \|_{\mc{L}_\infty}=    0.7413$ (pu), and $\| \m w \|_{\mc{L}_\infty}=  0.2564$ (pu) respectively for the 9-, 39-, and 57-bus networks.  \textcolor{black}{The convergence of the SCA  to compute a centralized $\m K$ on the 57-bus network is demonstrated in Fig.~\ref{Fig:Hmu} as an example. Similar convergence plots for the decentralized or input-constrained controllers are obtained.  In Fig.~\ref{Fig:Hmu}, the value of $\hbar$ represents the optimal objective value of problem~\eqref{equ:LINF-Niko} per iteration. The quantity $\mu^2$ is the expression $\mu_0\mu_1+\mu_2$  obtained from the optimal values of variables $\mu_0$, $\mu_1$, and $\mu_2$ in problem~\eqref{equ:LINF-Niko} per iteration.}

Table~\ref{Table:CentralizedDecentralizedSummary} summarizes the performance of the centralized, input-constrained, and decentralized controllers when computed by $\mathrm{MaxIter}=50$ iterations of their respective SCA Algorithm~\ref{algorithm:SCA}.  \textcolor{black}{For instance, notice the squared root of the value at the last iteration in Fig.~\ref{Fig:Hmu} equals $1.1621$ which is listed in Column 3 of Table~\ref{Table:CentralizedDecentralizedSummary} corresponding to the centralized controller for the 57-bus system. }
The trend is that the centralized and the input-constrained achieve similar performances.   The decentralized $\mc{L}_{\infty}$ control also performs well in terms of curbing frequency and voltage deviations. On the other hand, the decentralized controller shows poorer performance  in controlling the voltage of the 39-bus system. \textcolor{black}{In this case, only local state deviations are used to compute the feedback gain}.   

\begin{figure}[t]
\color{black}
	\centering
		\subfloat[]{\includegraphics[scale=0.24]{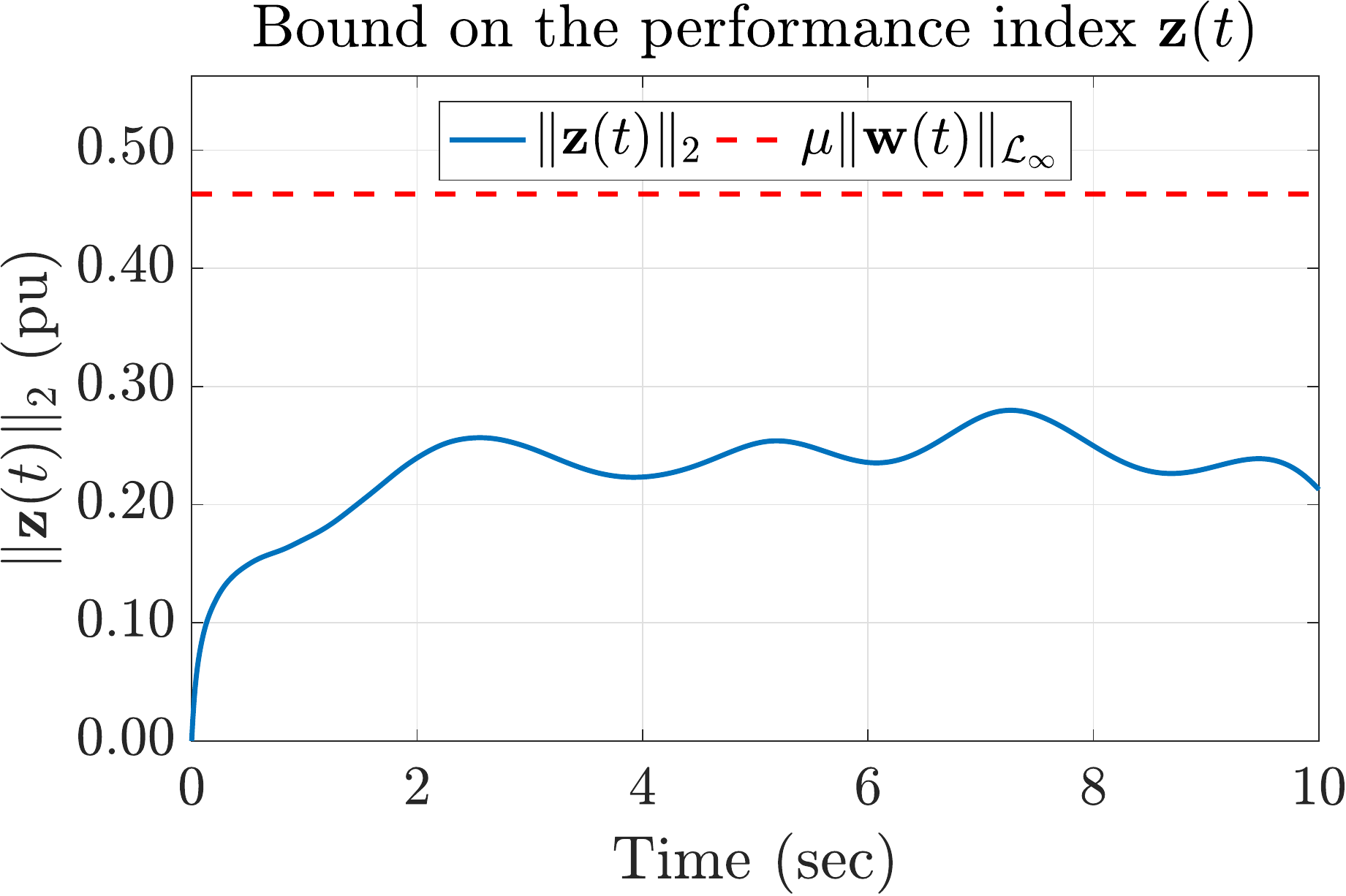}\label{Fig:PB39}}  
				\subfloat[]{\includegraphics[scale=0.24]{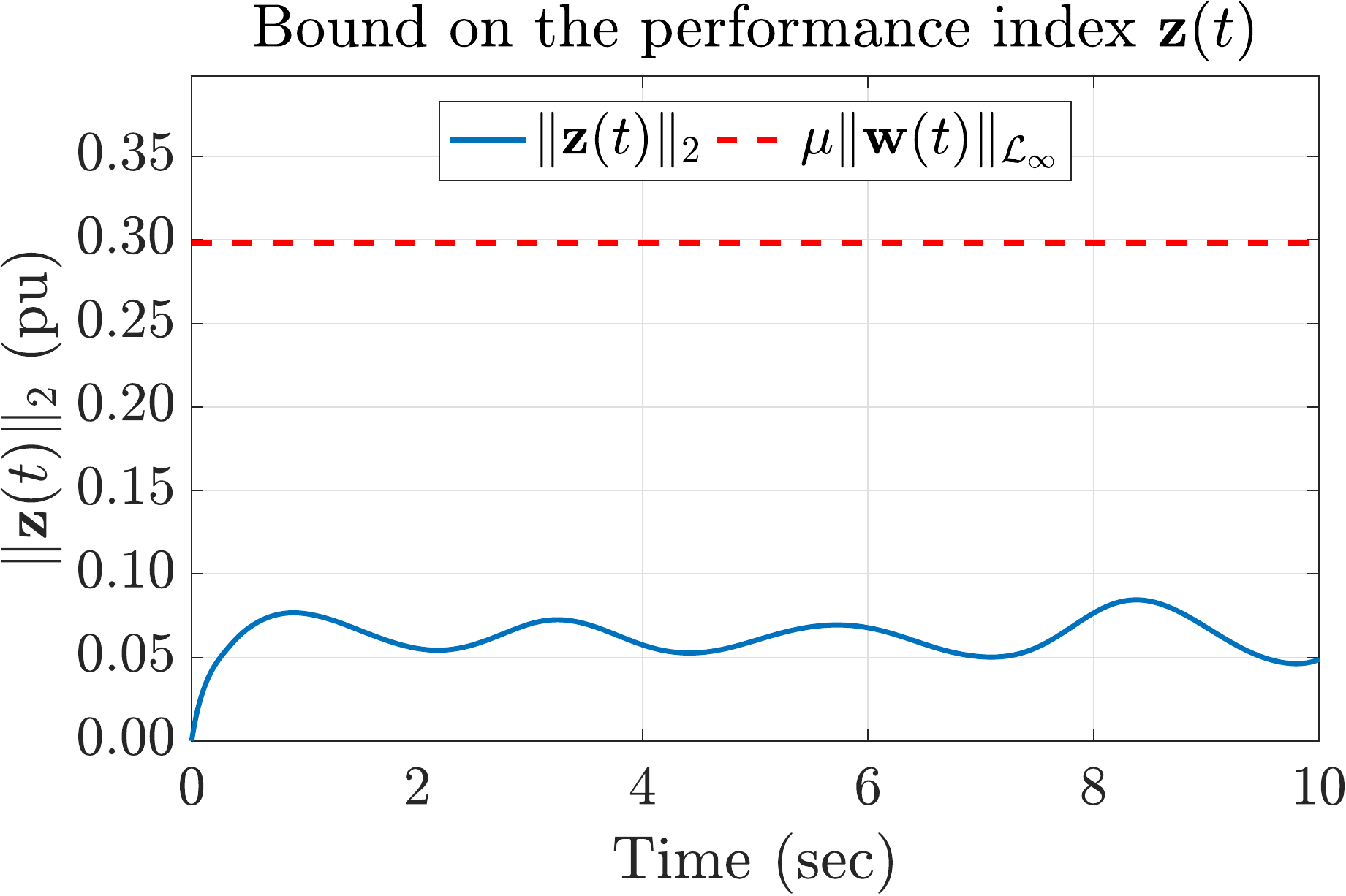}\label{Fig:PB57}} 
	\caption{Bound on the performance index  for the centralized controller on   \protect\subref{Fig:PB39} the 39-bus network, and \protect\subref{Fig:PB57} the 57-bus network.   \label{Fig:PB}}
\end{figure}

Finally, we further bring to attention Columns 3 and 4 from Table~\ref{Table:CentralizedDecentralizedSummary} where it always holds that $\mu \le  \frac{\|\m\Delta\m z\|_{\mc{L}_\infty}}{\|\m\Delta \m w\|_{\mc{L}_\infty}}$. \textcolor{black}{To emphasize this result,  corresponding plots for the 39-bus and 57-bus networks for the centralized controller are also provided in Fig.~\ref{Fig:PB}}.  In short, the results shown in Fig.~\ref{Fig:PB} and Table~\ref{Table:CentralizedDecentralizedSummary} corroborate the findings of Theorem~\ref{th1} that the performance bound for the $\mc{L}_{\infty}$ controller will be satisfied for the power network under various control architectures (centralized, decentralized, and input-constrained). This illustrates that the result of Theorem~\ref{th1} is not merely theoretical---it provides a useful way of bounding the performance of the system in terms of the worst-case disturbance while not yielding too conservative results. It can be seen that the performance level $\mu ||\m w(t)||_{\mathcal{L}_{\infty}}$ is not significantly larger than $||\m z(t)||_2$.


\begin{table}[t]
\centering
\color{black}
	\scriptsize
	\renewcommand{\arraystretch}{1}
	\caption{\small Performance of the $\mathcal{L}_{\infty}$ controller for different test cases, under various control architectures (centralized, input-constrained, and decentralized).}
	\begin{tabular}{|c|c|c|c|c|c|c|c|c|}
		\hline
		Case         & Type            & $\mu$           & $\frac{\|\m\Delta\m z\|_{\mc{L}_\infty}}{\|\m\Delta \m w\|_{\mc{L}_\infty}}$ & \makecell[c]{Freq.\\ Dev. (Hz)} & \makecell[c]{ Volt.\\ Dev. (pu)}   & \makecell[c]{ $\|\m\Delta \m u\|_{\mc{L}_{\infty}}$ \\   (pu)}   \\
		\hline \hline
		\multirow{3}{*}{9-bus}   & Cen.                 & 0.9573       &      0.4159
		           &       0.0029      &  0.0130
		       &        0.1275 \\      
		& Inp.                   &  0.9445         &        0.4372             &    0.0025        &    0.0092            &         0.1312 \\    
		& Dec.                 &  9.8712         &     1.1416            &   0.0053        &    0.0186
		     &   0.3736 \\            
		\hline \hline
		\multirow{3}{*}{39-bus} & Cen.                    &   0.6241        & 0.3775              &  0.0183
		      &0.0386
		             &    1.10561 \\        
		& Inp.                 &  0.8654     &  0.3211         & 0.0175
		        & 0.0422
		                     &        0.9816
		                     	 \\     
		& Dec.                &  3.4654      &     0.8870           &         0.0310  &    0.1075         &  2.8432    \\ 
		\hline \hline
		\multirow{3}{*}{57-bus}        & Cen.         &    1.1621
		         &        0.3293
		             &     0.0076      &   0.0155 &    0.2988 \\ 
		& Inp.                   &    2.5334        &    0.3087
		                  &     0.0074
		                       & 0.0163
		                             &     0.2762\\
		& Dec.                  & 5.2741         &     1.2271
		           &     0.0124
		              &      0.0323     &     1.1601
		              \\  
		\hline
	\end{tabular}
	\label{Table:CentralizedDecentralizedSummary}
	\vspace{-0.5cm}
\end{table}
\color{black}

\color{black}
\subsection{$\mc{L}_{\infty}$ control under large non-zero mean wind disturbances}
\begin{figure}[t]
	\subfloat[]{\includegraphics[scale=0.23]{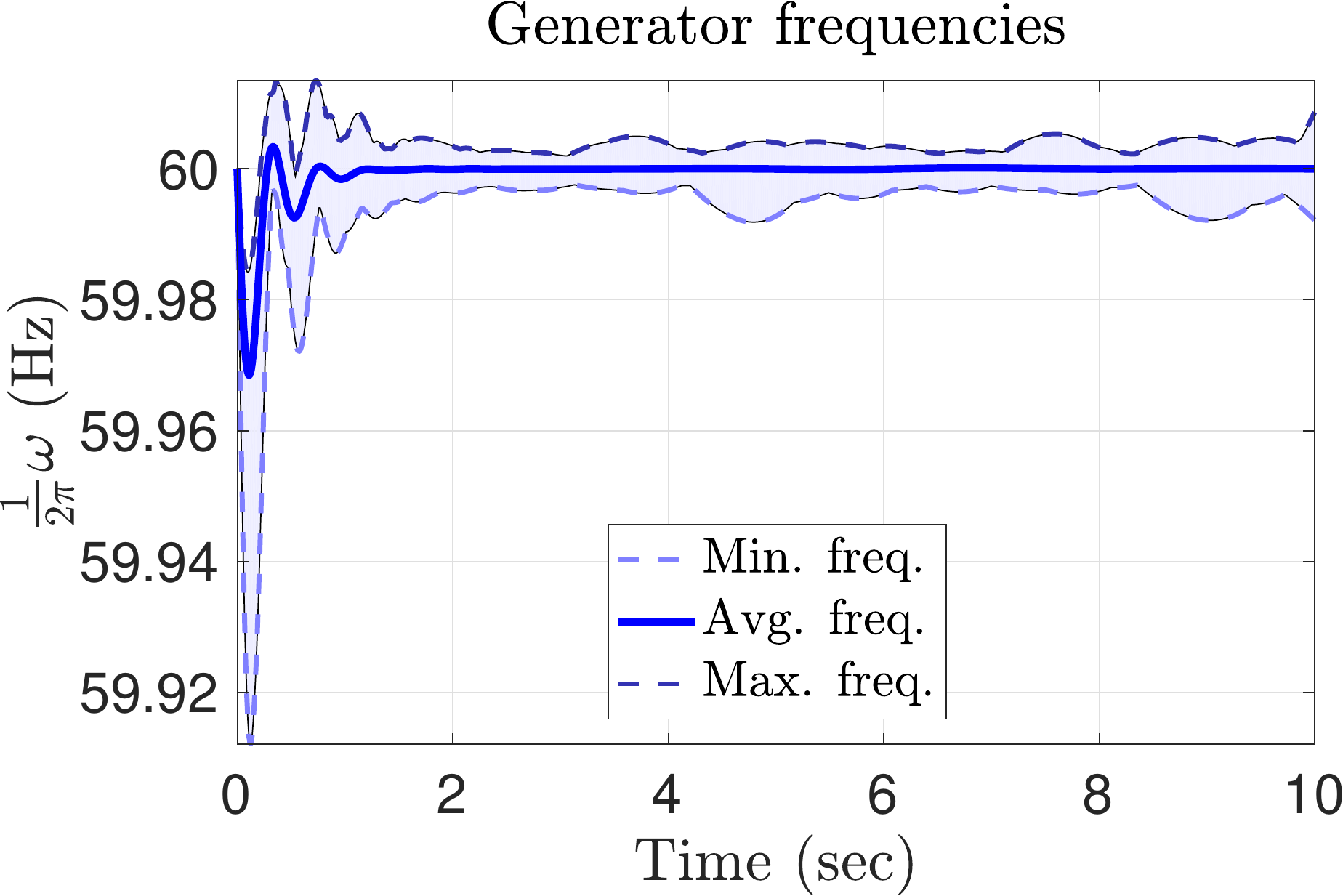}\label{Fig:CentFreqLarge}} 
	\subfloat[]{\includegraphics[scale=0.23]{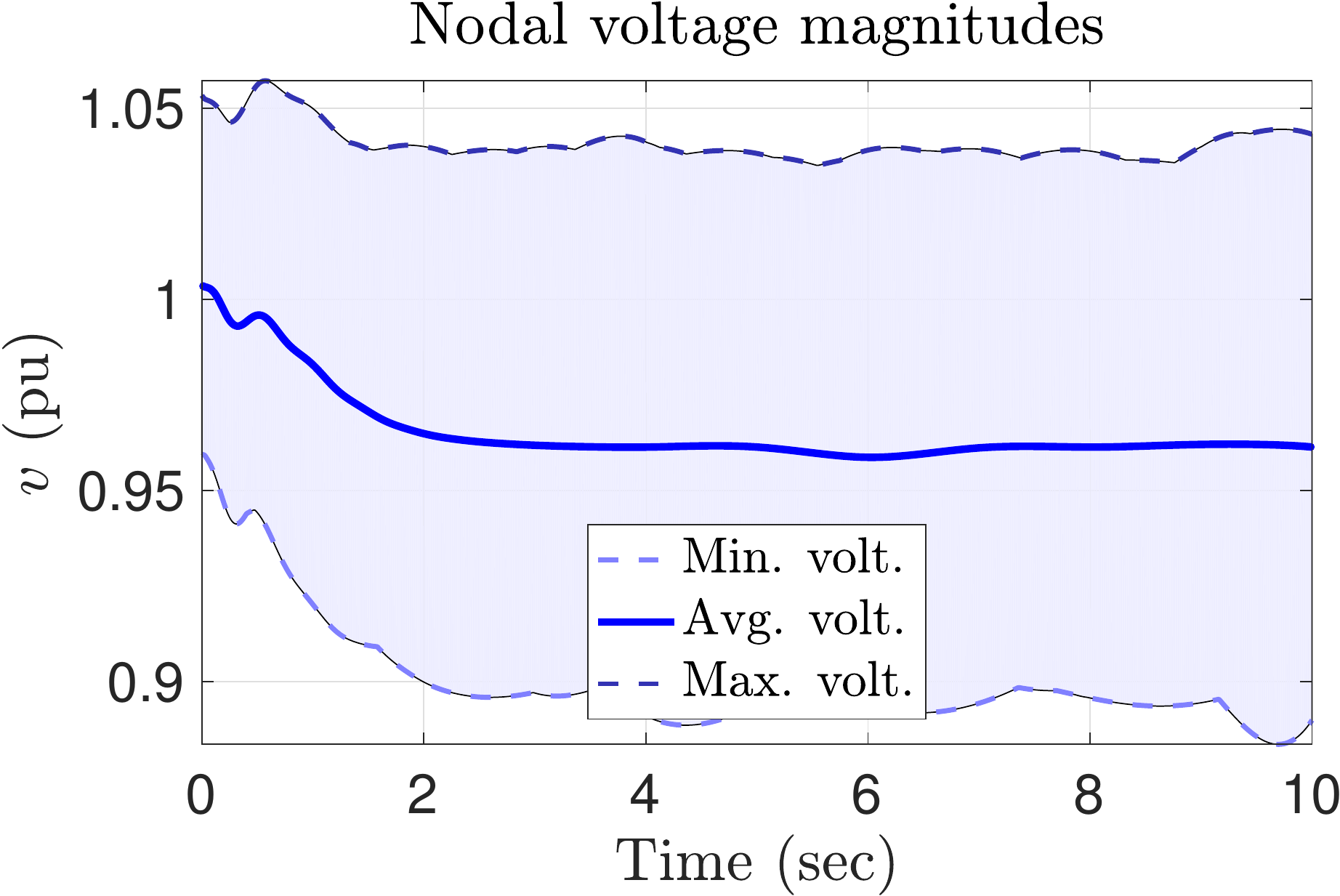}\label{Fig:CentVMagsLarge}} 
	\caption{	\textcolor{black}{Range and averages of frequency and voltage plots for 50  large wind and load disturbances and across all buses.} \label{Fig:LargeDisturbance}}
	\vspace{-0.55cm}
\end{figure}

In this setup,  the disturbance model~\eqref{EqGroup:NoiseVariations} is modified to include random, large step disturbances in  wind generation as well. The new disturbance model is as follows:
\begin{IEEEeqnarray}{rCl}
	\IEEEyesnumber \label{EqGroup:StepNoiseVariations}
	p_{l_n}(t)&=& p_{l_n}^0+ \Delta p_{l_n}+z_{l_n}(t), n \in \mc{N} \IEEEyessubnumber \label{Eq:TimeVaryingpl2}  \\
	p_{r_n}(t)&=& p_{r_n}^0-\Delta p_{r_n}+z_{r_n}(t), n \in \mc{R} \IEEEyessubnumber \label{Eq:TimeVaryingpr2}
\end{IEEEeqnarray}
where  quantities $z_{l_n}(t)$ and $z_{r_n}(t)$ are similar to~\eqref{EqGroup:NoiseVariations}.  However, the quantity $\Delta p_{r_n}^0$ is included as a step disturbance in wind generation. Its value is random and allowed to vary in the interval $[0,p_{r_n}^0]$ simulating a sudden loss of wind generation of up to $100\%$ in seconds.   The simulations are conducted for 50 such random realizations.  The frequency and voltage deviations  with respect to the initial point $\m x^0$ and across all 50  realizations are recorded.   Ranges and averages of frequency and voltage plots are provided in Fig.~\ref{Fig:LargeDisturbance}. The results show that even under significant, unpredictable changes in wind generation, the $\mathcal{L}_{\infty}$ still ensures frequency and voltage stability.  We note that AGC and $\mc{H}_{\infty}$ both failed in producing bounded state trajectories under large wind disturbances, as the used DAE solver diverged. 

\color{black}
\vspace{-0.255cm}
\section{Concluding Remarks and Future Work}
The paper considers a new notion of robust feedback control in power networks, namely the $\mathcal{L}_{\infty}$ control which considers worst-case bounds on uncertainty from renewables and loads while accounting for input bound constraints and various control, centralized/decentralized architectures. The proposed robust controller is applied on the power network dynamics with nonlinear DAE models under significant load and renewables uncertainty. The performance of the controller shows improvement over the status-quo controllers that use similar information to $\mathcal{L}_{\infty}$ control. The paper's limitations are two-fold. First, the proposed controller requires the knowledge of the grid operating point---and is not robust to changes in the operating points unless these changes are incorporated in the controller synthesis through the state-space matrices and Theorem~\ref{th1}. Second, and similar to recent feedback control methods in the literature, it is still unclear why the linear controller based on the linearized grid model works well for the nonlinear DAE grid model. This paper does not provide answers to the aforementioned limitations. 

Future work will focus on the following topics: \textit{(i)} Deriving $\mathcal{L}_{\infty}$ stability conditions of the nonlinear DAEs and then a corresponding controller which does not require the knowledge of the grid's operating point. \textit{(ii)} Scaling the proposed controller using SDP solvers that exploit sparsity of the state-space matrices. \textit{(iii)} Investigating whether $\mathcal{L}_{\infty}$ controller---a robust control method---yields a cheaper or more expensive overall power network operational costs, in comparison with AGC and other control architectures. 

\normalcolor
\bibliographystyle{IEEEtran}
\bibliography{mybib}

\begin{thebibliography}{10}
\providecommand{\url}[1]{#1}
\csname url@samestyle\endcsname
\providecommand{\newblock}{\relax}
\providecommand{\bibinfo}[2]{#2}
\providecommand{\BIBentrySTDinterwordspacing}{\spaceskip=0pt\relax}
\providecommand{\BIBentryALTinterwordstretchfactor}{4}
\providecommand{\BIBentryALTinterwordspacing}{\spaceskip=\fontdimen2\font plus
\BIBentryALTinterwordstretchfactor\fontdimen3\font minus
  \fontdimen4\font\relax}
\providecommand{\BIBforeignlanguage}[2]{{%
\expandafter\ifx\csname l@#1\endcsname\relax
\typeout{** WARNING: IEEEtran.bst: No hyphenation pattern has been}%
\typeout{** loaded for the language `#1'. Using the pattern for}%
\typeout{** the default language instead.}%
\else
\language=\csname l@#1\endcsname
\fi
#2}}
\providecommand{\BIBdecl}{\relax}
\BIBdecl

\bibitem{lei2009review}
M.~Lei, L.~Shiyan, J.~Chuanwen, L.~Hongling, and Z.~Yan, ``A review on the
  forecasting of wind speed and generated power,'' \emph{Renewable and
  Sustainable Energy Reviews}, vol.~13, no.~4, pp. 915--920, 2009.

\bibitem{MalladaTang2013}
E.~Mallada and A.~Tang, ``Dynamics-aware optimal power flow,'' in \emph{Proc.
  52nd IEEE Conf. Decision and Control}, Dec. 2013, pp. 1646--1652.

\bibitem{bazrafshan2017coupling}
M.~Bazrafshan, N.~Gatsis, A.~Taha, and J.~A. Taylor, ``Coupling load-following
  stability with opf,'' \emph{IEEE Transactions on Smart Grid}, 2017.

\bibitem{siano2014demand}
P.~Siano, ``Demand response and smart grids—a survey,'' \emph{Renewable and
  Sustainable Energy Reviews}, vol.~30, pp. 461--478, 2014.

\bibitem{Nguyen2016}
H.~D. Nguyen and K.~Turitsyn, ``{Robust Stability Assessment in the Presence of
  Load Dynamics Uncertainty},'' \emph{IEEE Trans. Power Syst.}, vol.~31, no.~2,
  pp. 1579--1594, Mar. 2016.

\bibitem{Vu2017}
T.~L. Vu and K.~Turitsyn, ``{A Framework for Robust Assessment of Power Grid
  Stability and Resiliency},'' \emph{IEEE Trans. Autom. Control}, vol.~62,
  no.~3, pp. 1165--1177, Mar. 2017.

\bibitem{Siljak2002}
D.~Siljak, D.~Stipanovic, and A.~Zecevic, ``{Robust decentralized
  turbine/governor control using linear matrix inequalities},'' \emph{IEEE
  Trans. Power Syst.}, vol.~17, no.~3, pp. 715--722, Aug. 2002.

\bibitem{Marinovici2013}
L.~D. Marinovici, J.~Lian, K.~Kalsi, P.~Du, and M.~Elizondo, ``{Distributed
  Hierarchical Control Architecture for Transient Dynamics Improvement in Power
  Systems},'' \emph{IEEE Trans. Power Syst.}, vol.~28, no.~3, pp. 3065--3074,
  Aug. 2013.

\bibitem{Lian2017}
\BIBentryALTinterwordspacing
J.~Lian, S.~Wang, R.~Diao, and Z.~Huang, ``{Decentralized Robust Control for
  Damping Inter-area Oscillations in Power Systems},'' 2017. [Online].
  Available: \url{https://arxiv.org/pdf/1701.02036.pdf}
\BIBentrySTDinterwordspacing

\bibitem{Fosha1970}
C.~Fosha and O.~Elgerd, ``{The Megawatt-Frequency Control Problem: A New
  Approach Via Optimal Control Theory},'' \emph{IEEE Trans. Power App. Syst.},
  vol. PAS-89, no.~4, pp. 563--577, Apr. 1970.

\bibitem{Lin2011}
F.~Lin, M.~Fardad, and M.~R. Jovanovic, ``{Augmented Lagrangian Approach to
  Design of Structured Optimal State Feedback Gains},'' \emph{IEEE Trans.
  Autom. Control}, vol.~56, no.~12, pp. 2923--2929, Dec. 2011.

\bibitem{Naguru2015}
N.~R. Naguru and V.~Sarkar, ``{Optimal wide area control of a power system with
  limited measurements},'' in \emph{2015 IEEE International Conference on
  Signal Processing, Informatics, Communication and Energy Systems
  (SPICES)}.\hskip 1em plus 0.5em minus 0.4em\relax IEEE, Feb. 2015, pp. 1--5.

\bibitem{LianChakrabortty2017}
\BIBentryALTinterwordspacing
F.~Lian, A.~Chakrabortty, and A.~Duel-Hallen, ``{Game-Theoretic Multi-Agent
  Control and Network Cost Allocation Under Communication Constraints},''
  \emph{IEEE J. Selected Areas in Comm.}, vol.~35, no.~2, pp. 330--340, Feb.
  2017. [Online]. Available: \url{http://ieeexplore.ieee.org/document/7835692/}
\BIBentrySTDinterwordspacing

\bibitem{CandesBoyd2008}
E.~J. Cand{\`{e}}s, M.~B. Wakin, and S.~P. Boyd, ``{Enhancing Sparsity by
  Reweighted l-1 Minimization},'' \emph{Journal of Fourier Analysis and
  Applications}, vol.~14, no. 5-6, pp. 877--905, Dec. 2008.

\bibitem{Lin2013}
F.~Lin, M.~Fardad, and M.~R. Jovanovic, ``{Design of Optimal Sparse Feedback
  Gains via the Alternating Direction Method of Multipliers},'' \emph{IEEE
  Trans. Autom. Control}, vol.~58, no.~9, pp. 2426--2431, Sept. 2013.

\bibitem{Dorfler2014}
F.~Dorfler, M.~R. Jovanovic, M.~Chertkov, and F.~Bullo, ``{Sparsity-Promoting
  Optimal Wide-Area Control of Power Networks},'' \emph{IEEE Trans. Power
  Syst.}, vol.~29, no.~5, pp. 2281--2291, Sept. 2014.

\bibitem{Wu2016a}
X.~Wu, F.~Dorfler, and M.~R. Jovanovic, ``{Input-Output Analysis and
  Decentralized Optimal Control of Inter-Area Oscillations in Power Systems},''
  \emph{IEEE Trans. Power Syst.}, vol.~31, no.~3, pp. 2434--2444, May 2016.

\bibitem{PiroozAzad2016}
S.~{Pirooz Azad}, J.~A. Taylor, and R.~Iravani, ``{Decentralized Supplementary
  Control of Multiple LCC-HVDC Links},'' \emph{IEEE Trans. Power Syst.},
  vol.~31, no.~1, pp. 572--580, Jan. 2016.

\bibitem{Wytock2013}
\BIBentryALTinterwordspacing
M.~Wytock and J.~Z. Kolter, ``{A Fast Algorithm for Sparse Controller
  Design},'' Dec. 2013. [Online]. Available:
  \url{http://arxiv.org/abs/1312.4892}
\BIBentrySTDinterwordspacing

\bibitem{Singh2016}
A.~K. Singh and B.~C. Pal, ``{Decentralized Control of Oscillatory Dynamics in
  Power Systems Using an Extended LQR},'' \emph{IEEE Trans. Power Syst.},
  vol.~31, no.~3, pp. 1715--1728, May 2016.

\bibitem{Singh2017}
------, ``{Decentralized Nonlinear Control for Power Systems using Normal Forms
  and Detailed Models},'' \emph{IEEE Transactions on Power Systems}, 2017, to
  be published.

\bibitem{Schuler2014a}
S.~Schuler, U.~M{\"{u}}nz, and F.~Allg{\"{o}}wer, ``{Decentralized state
  feedback control for interconnected systems with application to power
  systems},'' \emph{Journal of Process Control}, vol.~24, no.~2, pp. 379--388,
  Feb. 2014.

\bibitem{Schaab2017}
K.~Schaab, J.~Hahn, M.~Wolkov, and O.~Stursberg, ``{Robust control for voltage
  and transient stability of power grids relying on wind power},''
  \emph{Control Engineering Practice}, vol.~60, pp. 7--17, Mar. 2017.

\bibitem{Bevrani2016}
H.~Bevrani, M.~R. Feizi, and S.~Ataee, ``Robust frequency control in an
  islanded microgrid: ${H} _{\infty}$ and $\mu$-synthesis approaches,''
  \emph{IEEE Transactions on Smart Grid}, vol.~7, no.~2, pp. 706--717, March
  2016.

\bibitem{pancake2000d}
T.~A. Pancake, ``Analysis and control of uncertain/nonlinear systems in the
  presence of bounded disturbance inputs,'' Ph.D. dissertation, Purdue
  University, 2000.

\bibitem{sauerpai1998}
P.~W. Sauer and M.~A. Pai, \emph{Power System Dynamics and Stability}.\hskip
  1em plus 0.5em minus 0.4em\relax Prentice Hall, 1998.

\bibitem{CAISO}
\BIBentryALTinterwordspacing
CAISO, ``Real time demand curves,'' July 2017. [Online]. Available:
  \url{http://www.caiso.com/outlook/SystemStatus.html}
\BIBentrySTDinterwordspacing

\bibitem{Dinh2012}
Q.~T. Dinh, S.~Gumussoy, W.~Michiels, and M.~Diehl, ``Combining convex concave
  decompositions and linearization approaches for solving bmis, with
  application to static output feedback,'' \emph{IEEE Transactions on Automatic
  Control}, vol.~57, no.~6, pp. 1377--1390, June 2012.

\bibitem{lofberg2004yalmip}
J.~Lofberg, ``Yalmip: A toolbox for modeling and optimization in matlab,'' in
  \emph{Computer Aided Control Systems Design, 2004 IEEE International
  Symposium on}.\hskip 1em plus 0.5em minus 0.4em\relax IEEE, 2004, pp.
  284--289.

\bibitem{mosek2015mosek}
A.~Mosek, ``The mosek optimization toolbox for matlab manual,'' \emph{Version
  7.1 (Revision 28)}, p.~17, 2015.

\bibitem{zimmerman2011matpower}
R.~D. Zimmerman, C.~E. Murillo-S{\'a}nchez, and R.~J. Thomas, ``Matpower:
  Steady-state operations, planning, and analysis tools for power systems
  research and education,'' \emph{IEEE Transactions on power systems}, vol.~26,
  no.~1, pp. 12--19, 2011.

\bibitem{sauerpower}
P.~W. Sauer, M.~Pai, and J.~H. Chow, ``Power system toolbox,'' \emph{Power
  System Dynamics and Stability: With Synchrophasor Measurement and Power
  System Toolbox 2e: With Synchrophasor Measurement and Power System Toolbox},
  pp. 305--325.

\bibitem{ZhaoTopcuLiLow2014}
C.~Zhao, U.~Topcu, N.~Li, and S.~Low, ``Design and stability of load-side
  primary frequency control in power systems,'' \emph{IEEE Trans. Autom.
  Control}, vol.~59, no.~5, pp. 1177--1189, May 2014.

\bibitem{gahinet1994linear}
P.~Gahinet and P.~Apkarian, ``A linear matrix inequality approach to h∞
  control,'' \emph{International journal of robust and nonlinear control},
  vol.~4, no.~4, pp. 421--448, 1994.

\bibitem{LeXie2011}
{Le Xie}, P.~M.~S. Carvalho, L.~A. F.~M. Ferreira, {Juhua Liu}, B.~H. Krogh,
  N.~Popli, and M.~D. Ili{\'{c}}, ``{Wind Integration in Power Systems:
  Operational Challenges and Possible Solutions},'' \emph{Proc. IEEE}, vol.~99,
  no.~1, pp. 214--232, Jan 2011.

\bibitem{wollenbergbook2012}
A.~J. Wood and B.~F. Wollenberg, \emph{Power Generation, Operation, and
  Control}, 3rd~ed.\hskip 1em plus 0.5em minus 0.4em\relax John Wiley \& Sons,
  2012.

\bibitem{WangLiuLowMei2017}
\BIBentryALTinterwordspacing
Z.~Wang, F.~Liu, J.~Z.~F. Pang, S.~Low, and S.~Mei, ``{Distributed Optimal
  Frequency Control Considering a Nonlinear Network-Preserving Model},'' 2017.
  [Online]. Available: \url{https://arxiv.org/pdf/1709.01543.pdf}
\BIBentrySTDinterwordspacing

\end{thebibliography}

\appendices

\section{Proof of Theorem~\ref{th1}}~\label{proofth1}
To prove Theorem~\ref{th1}, the following lemma is used.
\begin{mylem}[From~\cite{pancake2000d}]~\label{lem:l1}
	Consider a quadratic Lyapunov function $V(\m x(t)):= \m x^{\top}(t)\m P\m x(t)$. Suppose there exists $\m P \succ \m O$, and scalars $\{\mu_0,\mu_1,\mu_2\}>0$ such that for all $\m x(t)$ and $\m w(t)$ we have
	$$\dot{V}(\m x(t)) <0 \;\; \text{when} \; \; \m x^{\top}(t)\m P\m x(t) > \mu_0 \Vert \m w(t) \Vert^2_2, \; \text{and}$$  
	$$\Vert \m z(t) \Vert_2^2  \leq \mu_1 \m x^{\top}(t)\m P\m x(t) +\mu_2 \Vert \m w(t) \Vert^2_{2}.$$
	Then the closed loop system with unknown inputs~\eqref{equ:closedloop} is $\mc{L}_{\infty}$-stable with performance level $\mu=\sqrt{\mu_0\mu_1+\mu_2}$. 
\end{mylem} 
\begin{IEEEproof}[Proof of Theorem~\ref{th1}]
	Consider a quadratic Lyapunov function $V(\m x(t)):= \m x^{\top}(t)\m P\m x(t)$ where $\m P \succ \m O$. We consider classical conditions on the existence of this Lyapunov function from invoking the S-procedure  in Lemma~\ref{lem:l1}
	$$ \dot{V}(\m x(t)) \leq - \alpha (\m x^{\top}(t)\m P\m x(t)-\mu_0 \Vert \m w(t) \Vert ^2_2),$$
	for positive scalars $\alpha$ and $ \mu_0$. Substituting the time-derivative of $V(\m x)$ yields
	\begin{equation}\label{Lder}
	\dot{\m x}^{\top}(t)\m P\m x(t)+\m x^{\top}(t)\m P\dot{\m x}(t) \vspace{-0.15cm}
	\end{equation} 
	$$ + \alpha \m x^{\top}(t)\m P\m x(t) - \alpha \mu_0 \m w^{\top}(t)\m w(t) \leq 0.$$
	Substituting $\dot{\m x}(t) = (\m A+\m B_u \m K)\m x(t)+\m B_w \m w(t)$ in \eqref{Lder} yields
	\begin{align*}
	\begin{bmatrix}
	\m x(t)\\ \m w(t)
	\end{bmatrix}^{\top}
	\begin{bmatrix}
	(\m A^{\top}+\m K^{\top}\m B_u^{\top})\m P\\+\m P(\m A+\m B_u\m K)+\alpha \m P & \m P\m B_w \\
	\m B_w^{\top}\m P & -\alpha \mu_0\m I
	\end{bmatrix}
	\begin{bmatrix}
	\m x(t)\\ \m w(t)
	\end{bmatrix} \leq 0
	\end{align*}
	which is equivalent to
	\begin{align*}
	\begin{bmatrix}
	(\m A^{\top}+\m K^{\top}\m B_u^{\top})\m P+\m P(\m A+\m B_u\m K)+\alpha \m P & \m P\m B_w \\
	\m B_w^{\top}\m P & -\alpha \mu_0 \m I
	\end{bmatrix} \preceq 0.
	\end{align*}
	Applying congruence transformation with $\m S = \m P^{-1} \succ 0$ yields
	\begin{align*}
	\begin{bmatrix}
	\m S & \m O \\ \m O & \m I
	\end{bmatrix}
	&\begin{bmatrix}
	(\m A^{\top}+\m K^{\top}\m B_u^{\top})\m P\\+\m P(\m A+\m B_u\m K)+\alpha \m P & \m P\m B_w \\
	\m B_w^{\top}\m P & -\alpha \mu_0 \m I
	\end{bmatrix} 
	\begin{bmatrix}
	\m S & \m O \\ \m  O & \m I
	\end{bmatrix}
	\preceq 0,
	\end{align*}
	which can be written as
	\begin{equation}~\label{LMI1}
	\hspace{-0.55cm}	\begin{bmatrix}
	\m S\m A^{\top}+\m A\m S+\m Z^{\top}\m B_u^{\top}+\m B_u\m Z+\alpha \m S & \m B_w \\
	\m B_w^{\top} & -\alpha \mu_0 \m I
	\end{bmatrix} \preceq 0, 
	\end{equation}
	where $\m Z$ and $\m S$ are the matrix variables and $\m K = \m Z \m S^{-1}$ is the feedback gain matrix. This verifies the first matrix inequality in~\eqref{equ:LINF}.
	From the second condition in Lemma~\ref{lem:l1}, it is required that
	\begin{equation}~\label{equ:zcond}
	\Vert \m z(t) \Vert_2^2 \leq \mu_1 V(\m x(t))+\mu_2 \Vert \m w(t) \Vert^2_{2}.
	\end{equation}
	for positive scalars $\mu_1$ and $\mu_2$. Given the definition of the performance index $\m z(t)$ and the candidate Lyapunov function, we obtain
	\begin{eqnarray*}
		\Vert \m C\m x(t) + \m D\m u(t)\Vert ^2_2 \leq \mu_1\m x^{\top}(t)\m P\m x(t)+\mu_2 \Vert \m w(t) \Vert ^2_2.
	\end{eqnarray*}
	Substituting $\m u(t) = \m K \m x(t)$ in the previous equation, we obtain
	\begin{align*}
	\m x^{\top}(t)\left( \left(\m C+\m D\m K\right)^{\top}\left(\m C+\m D\m K\right)-\mu_1 \m P \right)\m x(t) \\
	\hspace{0.5cm}- \mu_2 \m w^{\top}(t)\m w(t) \leq 0
	\end{align*}
	which can be written as
	\begin{equation*}
	\begin{bmatrix}
	\m x(t)\\ \m w(t)
	\end{bmatrix}^{\top}
	\begin{bmatrix}
	\left(\m C+\m D\m K\right)^{\top}\left(\m C+\m D\m K\right) \\ 
	- \mu_1\m P  & \m O\\ \m O & -\mu_2\m I
	\end{bmatrix}
	\begin{bmatrix}
	\m x(t)\\ \m w(t)
	\end{bmatrix} \leq 0.
	\end{equation*}
	Equivalently, we obtain
	$$\begin{bmatrix}
	-\mu_1\m P& \m O &(\m C+\m D\m K)^{\top}\\ \m O & -\mu_2\m I & \m O \\ \m C+\m D\m K& \m O & -\m I
	\end{bmatrix}
	\preceq 0 .$$
	Applying congruence transformation with $\m S = \m P^{-1} \succ 0$ yields
	\begin{align*}
	\begin{bmatrix}
	\m S & \m O &   \m O \\ \m O & \m I & \m O\\ \m O& \m O& \m I
	\end{bmatrix}
	&\cdot \begin{bmatrix}
	-\mu_1\m P& \m O &(\m C+\m D\m K)^{\top}\\ \m O & -\mu_2\m I & \m O \\ \m C+\m D\m K& \m O & -\m I
	\end{bmatrix}\\
	& 
	\cdot	\begin{bmatrix}
	\m S & \m O &   \m O \\ \m O & \m I & \m O\\ \m O& \m O& \m I
	\end{bmatrix}
	\preceq 0.
	\end{align*}
	Noticing that $\m K=\m Z\m S^{-1}$, we finally retrieve  
	\begin{equation}~\label{equ:secondMI}
	\begin{bmatrix}
	-\mu_1\m S & \m O & \m S\m C^{\top}+\m Z^{\top}\m D^{\top}\\ \m O & -\mu_2\m I & \m O \\ \m C\m S +\m D \m Z& \m O & -\m I
	\end{bmatrix}
	\preceq 0.
	\end{equation}
	This verifies the second matrix inequality in~\eqref{equ:LINF} and that the performance level is indeed $\mu=\sqrt{\mu_0\mu_1+\mu_2}$; see Lemma~\ref{lem:l1}. The third and fourth matrix inequalities in~\eqref{equ:LINF} guarantee Design Requirement 1. Lemma~\ref{lem:l1} is established based on the existence of an invariant ellipsoid as discussed in \cite{pancake2000d}. This invariant ellipsoid is described as 
	\begin{align}~\label{equ:invariantEllipsoid}
	\mc{E} = \lbrace \m x(t) \in \mathbb{R}^{n} | \m x(t)^{\top} \m P \m x(t) \leq \mu_0 \Vert \m w \Vert^2_{\mc{L}_\infty} \rbrace.
	\end{align}   
	Since $\mc{E}$ is invariant, then $\m x_0 \in \mc{E}$ guarantees $\m x(t) \in \mc{E}$ for all $t\geq t_0$. Suppose that the infinity norm of the disturbance signal is known, that is $\rho = \Vert \m w \Vert_{\mc{L}_\infty}$. This reflects the worst-case disturbance to the power network. Then, $\m x_0 \in \mc{E}$ implies
	\begin{align}
	\m x_0 \in \mc{E} &\Leftrightarrow \m x_0^{\top} \m P \m x_0 \leq \mu_0\rho^2 \nonumber \Leftrightarrow - \mu_0\rho^2+\m x_0^{\top} \m P \m x_0 \leq 0. \nonumber 
	\end{align}
	Applying Schur complement to the above equation and then substituting $\m S = \m P^{-1} \succ 0$
	establishes the third matrix inequality in~\eqref{equ:LINF} given by
	\begin{align}~\label{equ:invariantEllipsoid2}
	\bmat{-\mu_0\rho^2 & \m x_0^{\top} \\ \m x_0 & -\m S} \preceq 0.
	\end{align}
	To prove the fourth matrix inequality in~\eqref{equ:LINF}, substitute $\m u(t) = \m K \m x (t)$ with $ \m K = \m Z \m S^{-1}$. This yields
	\begin{eqnarray}
	&	\vert\vert \m u(t) \vert\vert_2^2 \leq \underset{t \geq t_0}{\mathrm{max}}\vert\vert \m u(t) \vert\vert_2^2 = \underset{t \geq t_0}{\mathrm{max}}\vert\vert \m Z \m S^{-1} \m x (t)\vert\vert_2^2 \nonumber \\
	\Rightarrow &	\underset{t \geq t_0}{\mathrm{max}}\vert\vert \m Z \m S^{-1} \m x (t)\vert\vert_2^2 = \vert\vert \m Z \m S^{-\frac{1}{2}} \vert\vert_2^2 \,\underset{t \geq t_0}{\mathrm{max}}\vert\vert S^{-\frac{1}{2}} \m x (t)\vert\vert_2^2. 	~\label{equ:inputConstraint1}
	\end{eqnarray}
	Assuming that~\eqref{equ:invariantEllipsoid2} is satisfied, then the following holds
	\begin{align}
	\vert\vert \m S^{-\frac{1}{2}} \m x (t)\vert\vert_2^2 = \m x (t)^{\top}  \m S^{-1} \m x (t) \leq \mu_0\rho^2 \nonumber
	\end{align}
	for all $t \geq t_0$, which consequently implies 
	\begin{align}
	\vert\vert S^{-\frac{1}{2}} \m x (t)\vert\vert_2^2 \leq \underset{t \geq t_0}{\mathrm{max}}\vert\vert S^{-\frac{1}{2}} \m x (t)\vert\vert_2^2 &\leq \mu_0\rho^2. \nonumber 
	\end{align}
	Based on this result,~\eqref{equ:inputConstraint1} can be written as
	\begin{align}
	\underset{t \geq t_0}{\mathrm{max}}\vert\vert \m Z \m S^{-1} \m x (t)\vert\vert_2^2 &\leq \vert\vert \m Z \m S^{-\frac{1}{2}} \vert\vert_2^2 \,\mu_0\rho^2 \nonumber \\ &\leq \lambda_{\max}(\m S^{-\frac{1}{2}} \m Z^{\top} \m Z \m S^{-\frac{1}{2}})\,\mu_0\rho^2.\label{equ:ineq}
	\end{align}
	If we upper bound the RHS of~\eqref{equ:ineq} with $u_{\max}^2$, then 
	\begin{align}~\label{equ:inputConstraint3}
	\lambda_{\max}(\m S^{-\frac{1}{2}} \m Z^{\top} \m Z \m S^{-\frac{1}{2}}) \,\mu_0\rho^2\leq u_{\max}^2 = \lambda_{\max}(u_{\max}^2\m I) 
	\end{align}
	such that
	$\vert\vert \m u(t) \vert\vert_2^2 \leq u_{\max}^2$
	which guarantees the input bound constraint. Notice that~\eqref{equ:inputConstraint3} implies
	$	-\frac{u_{\max}^2}{\rho^2}\m I+\mu_0\m S^{-\frac{1}{2}} \m Z^{\top} \m Z \m S^{-\frac{1}{2}} \preceq 0,$
	then applying the Schur complement and congruence transformation yields
	\vspace{-0.12cm}
	\begin{align}
	&\bmat{-\frac{u_{\max}^2}{\rho^2}\m I & \m S^{-\frac{1}{2}}\m Z^{\top}\\ \m Z \m S^{-\frac{1}{2}} & -\frac{1}{\mu_0} \m I} \preceq 0 \nonumber \\
	\Rightarrow	\begin{bmatrix}
	\m S^{\frac{1}{2}} & \m O \\ \m  O & \mu_0\m I
	\end{bmatrix}&\bmat{-\frac{u_{\max}^2}{\rho^2}\m I & \m S^{-\frac{1}{2}}\m Z^{\top}\\ \m Z \m S^{-\frac{1}{2}} & -\frac{1}{\mu_0} \m I} 	\begin{bmatrix}
	\m S^{\frac{1}{2}} & \m O \\ \m  O & \mu_0\m I
	\end{bmatrix}\preceq 0 \nonumber \\
	&\Rightarrow\bmat{-\frac{u_{\max}^2}{\rho^2}\m S & \mu_0\m Z^{\top}\\ \mu_0\m Z & -\mu_0 \m I} \preceq 0.
	\end{align}
	This completes the proof.
\end{IEEEproof}
\section{AGC Implementation}
\label{Appendix:AGC}
The AGC requires the addition of an extra state to the system, as follows \cite{wollenbergbook2012}:
\begin{IEEEeqnarray}{rCl}
	\label{EqGroup:yAGC}	\IEEEyesnumber \IEEEyessubnumber* 
	\dot{y}&=& K_{\mathrm{G}}\left(  -y - \mathrm{ACE}+ \sum\limits_{i \in \mc{G}} (p_{\mathrm{g}_i}-p_{\mathrm{g}_i}^0)\right) \label{Eq:yDot} \\
	r_i&=& r_i^0+K_{\mr{i}} y, i \in \mc{G} \label{Eq:prefParticipation}
\end{IEEEeqnarray}
where $K_{\mr{G}}$ is an integrator gain set to $1000$, $K_i=p_{\mr{g}_i}/\sum p_{\mr{g}_i}$ is the participation factor of each generator. Notice that the sum of participation factors equals to $1$, that is, $\sum_{i \in \mc{G}} K_i=1$.  Similar to \cite{WangLiuLowMei2017}, by treating the power network as a single control area, we select the following $\mr{ACE}$. 
\begin{IEEEeqnarray}{rCl}
	\mr{ACE}&=& \frac{1}{G}\sum\limits_{i \in \mc{G}}(\frac{1}{R}_i+D_i) (\omega_i - \omega^s). \label{Eq:ACE}
\end{IEEEeqnarray}
The control input to the exciter is computed according to the centralized $\mc{L}_{\infty}$ gain so that voltage control is not neglected.

\section{Proof of Theorem~\ref{th2}}~\label{proofth2}
\vspace{-0.54cm}
\begin{IEEEproof}[Proof of Theorem~\ref{th2}]
	Consider the first bilinear matrix inequality (BMI) constraint in~\eqref{equ:LINF}. Separating the bilinear terms in~\eqref{equ:LINF-BMI1} from the other terms yields
	\begin{align}
		~\label{equ:LINF-BMI1-1}
		\begin{bmatrix}
			\m S\m A^{\top}+\m A\m S\\ +\m Z^{\top}\m B_u^{\top}+\m B_u\m Z & \m B_w \\
			\m B_w^{\top} & \m O
		\end{bmatrix} + \alpha 
		\begin{bmatrix} 
			\m S & \m O \\ \m O & -\mu_0 \m I
		\end{bmatrix} \preceq 0 .
	\end{align}
	The bilinear term, which is the second term on the left-hand side of~\eqref{equ:LINF-BMI1-1}, is replaced by a convex function $\mc C_1(\alpha,\mu_0,\m S;\tilde{\alpha},\tilde{\mu}_0,\tilde{\m S})$ such that
	\begin{align}
		~\label{equ:LINF-BMI1-2}
		\alpha 
		\begin{bmatrix} 
			\m S & \m O \\ \m O & -\mu_0 \m I
		\end{bmatrix} = \alpha \m X \preceq \mc C_1(\alpha,\mu_0,\m S;\tilde{\alpha},\tilde{\mu}_0,\tilde{\m S}).
	\end{align}
	The expression $\alpha \m X$ can be written as
	\begin{align}
		~\label{equ:LINF-BMI1-4}
		\alpha \m X = \frac{1}{4}\big((\alpha \m I + \m X)^{\top}(\alpha \m I + \m X)  -(\alpha \m I - \m X)^{\top}(\alpha \m I - \m X)\big).
	\end{align}
	While $(\alpha \m I + \m X)^{\top}(\alpha \m I + \m X)$ is convex in $\alpha,\,\mu_0$, and $\m S$, the term $-(\alpha \m I - \m X)^{\top}(\alpha \m I - \m X)$ is concave. Since the first-order Taylor approximation of a concave function is indeed a global overestimator of the function, then the concave part of~\eqref{equ:LINF-BMI1-4} can be approximated by a linear function $F_{l}(\alpha,\mu_0,\m S;\tilde{\alpha},\tilde{\mu}_0,\tilde{\m S})$ that is calculated around $(\tilde{\alpha},\tilde{\mu}_0,\tilde{\m S})$. Consequently, for all $\tilde{\alpha},\tilde{\mu}_0,\tilde{\m S}$ and $\alpha,\mu_0,\m S$, we have
	\begin{align}
		~\label{equ:LINF-BMI1-5}
		-(\alpha \m I - \m X)^{\top}(\alpha \m I - \m X) \preceq F_{l}(\alpha,\mu_0,\m S;\tilde{\alpha},\tilde{\mu}_0,\tilde{\m S})
	\end{align}
	which implies that
	\begin{align}
		~\label{equ:LINF-BMI1-6}
		\mc C_1(\alpha,\mu_0,\m S;\tilde{\alpha},\tilde{\mu_0},\tilde{\m S}) = \frac{1}{4}(\alpha \m I + \m X)^{\top}(\alpha \m I + \m X) \nonumber \\ +  \frac{1}{4}F_{l}(\alpha,\mu_0,\m S;\tilde{\alpha},\tilde{\mu_0},\tilde{\m S}).
	\end{align}
	Let $\alpha = \tilde{\alpha} + \Delta\alpha$ and $\m X = \tilde{\m X} + \Delta \m X$ with $\Delta\alpha = \alpha - \tilde{\alpha}$, $\Delta \m X = \m X - \tilde{\m X}$, and
	\begin{align*}
		\tilde{\m X} = \begin{bmatrix} 
			\tilde{\m S} & \m O \\ \m O & -\tilde{\mu}_0 \m I
		\end{bmatrix}, \,\, \Delta \m X = \begin{bmatrix} 
			\m S - \tilde{\m S} & \m O \\ \m O & -(\mu_0-\tilde{\mu}_0) \m I
		\end{bmatrix}.
	\end{align*}
	Substituting these variables into the concave part of~\eqref{equ:LINF-BMI1-4} and removing all second-order terms that contain both $\Delta\alpha$ and $\Delta \m X$ produces
	\begin{align}
		~\label{equ:LINF-BMI1-7}
		F_{l}(\alpha,\mu_0,\m S;\tilde{\alpha},\tilde{\mu}_0,\tilde{\m S}) = -(\tilde{\alpha}^2+2\tilde{\alpha}\Delta\alpha)\m I+2\tilde{\alpha}\tilde{\m X} \nonumber \\ +2\tilde{\alpha}\Delta\m X+2\Delta\alpha\tilde{\m X}-\tilde{\m X}^2-\tilde{\m X}\Delta\m X -\Delta\m X\tilde{\m X}.
	\end{align}
	$F_{l}(\alpha,\mu_0,\m S;\tilde{\alpha},\tilde{\mu}_0,\tilde{\m S})$ in~\eqref{equ:LINF-BMI1-7} can be further expressed as
	\begin{align}
		\small \label{equ:LINF-BMI1-8}
		\hspace{-0.65cm}	F_{l}(\alpha,\mu_0,\m S;\tilde{\alpha},\tilde{\mu}_0,\tilde{\m S}) = \begin{bmatrix}
			F_{l_1}(\alpha,\m S;\tilde{\alpha},\tilde{\m S}) & \m O \\
			\m O & F_{l_2}(\alpha,\mu_0;\tilde{\alpha},\tilde{\mu}_0)
		\end{bmatrix} 
	\end{align}
	\begin{align}
		~\label{equ:LINF-BMI1-9}
		F_{l_1}(\alpha,\m S;\tilde{\alpha},\tilde{\m S}) = \,&\tilde{\alpha}^2\m I-2\tilde{\alpha}\tilde{\m S}+\tilde{\m S}^2\nonumber-2\alpha(\tilde{\alpha}\m I-\tilde{\m S}) \\ &+(2\tilde{\alpha}\m I-\tilde{\m S})\m S-\m S\tilde{\m S}\\
		F_{l_2}(\alpha,\mu_0;\tilde{\alpha},\tilde{\mu}_0) = &(\tilde{\alpha}^2+\tilde{\mu_0}^2+2\tilde{\alpha}\tilde{\mu}_0-2\alpha(\tilde{\alpha}+\tilde{\mu}_0)\nonumber \\ &-2\mu_0(\tilde{\alpha}+\tilde{\mu}_0))\m I.	~\label{equ:LINF-BMI1-10}
	\end{align}
	Thus, from~\eqref{equ:LINF-BMI1-2}--\eqref{equ:LINF-BMI1-6} and~\eqref{equ:LINF-BMI1-8}, the BMI in~\eqref{equ:LINF-BMI1} can be replaced by 
	\begin{align}
		~\label{equ:LINF-BMI1-11}
		\begin{bmatrix}
			\m S\m A^{\top}+\m A\m S+\m Z^{\top}\m B_u^{\top}\\+\m B_u\m Z+\frac{1}{4}F_{l_1}(\alpha,\m S;\tilde{\alpha},\tilde{\m S}) & \m B_w \\
			\m B_w^{\top} & \frac{1}{4}F_{l_2}(\alpha,\mu_0;\tilde{\alpha},\tilde{\mu}_0)
		\end{bmatrix} \nonumber\\
		+ \frac{1}{4}(\alpha \m I + \m X)^{\top}(\alpha \m I + \m X) \preceq 0.
	\end{align}
	Since 
	\begin{align*}
		\frac{1}{2}(\alpha \m I + \m X) = \begin{bmatrix}
			\frac{1}{2}(\alpha\m I+\m S) & \m O \\ \m O & \frac{1}{2}(\alpha-\mu_0)\m I
		\end{bmatrix},
	\end{align*}
	then applying Schur complement to~\eqref{equ:LINF-BMI1-11} yields
	\begin{align}
		\hspace{-0.85cm}	~\label{equ:LINF-BMI1-12}
		\begin{bmatrix}
			\m S\m A^{\top}+\m A\m S \\ +\m Z^{\top}\m B_u^{\top}+\m B_u\m Z & \star & \star & \star \\+\frac{1}{4}F_{l_1}(\alpha,\m S;\tilde{\alpha},\tilde{\m S}) \\ 
			\m B_w^{\top} & \frac{1}{4}F_{l_2}(\alpha,\mu_0;\tilde{\alpha},\tilde{\mu}_0) & \star & \star \\
			\frac{1}{2}(\alpha\m I+\m S) & \m O & -\m I & \star \\
			\m O & \frac{1}{2}(\alpha-\mu_0)\m I & \m O & -\m I 
		\end{bmatrix} \preceq 0
	\end{align}
	which is an LMI in $\alpha,\,\mu_0$, and $\m S$. This proves the convex approximation of the first constraint. 
	The other BMI constraints in~\eqref{equ:LINF} can be approximated with LMIs by applying a similar procedure based on the SCA. 
	For brevity, we do not provide these derivations here but still provide the linear form of the matrices the LMIs are given in this optimization problem 
	\begin{subequations}\label{equ:LINF-NikoP}
		\begin{align}
			\min& \;\;\;\hslash \\
			\st & \notag \\
			&\hspace{-0.6cm}	\begin{bmatrix}
				\m \Omega & \star & \star & \star \\
				\m B_w^{\top} & \frac{1}{4}F_{l_2}(\cdot) & \star & \star \\
				\frac{1}{2}(\alpha\m I+\m S) & \m O & -\m I & \star \\
				\m O & \frac{1}{2}(\alpha-\mu_0)\m I & \m O & -\m I 
			\end{bmatrix} \preceq 0 ~\label{equ:LINF-NikoP-LMI1}\\
			&\hspace{-0.6cm} \begin{bmatrix}
				\frac{1}{4}G_{l}(\cdot) & \star & \star & \star\\ \m O & -\mu_2\m I & \star & \star\\ \m C\m S +\m D \m Z& \m O & -\m I & \star \\ \frac{1}{2}(\mu_1\m I - \m S) & \m O & \m O & -\m I
			\end{bmatrix} \preceq 0 ~\label{equ:LINF-NikoP-LMI2}\\
			&\hspace{-0.6cm}\bmat{-\mu_0\rho^2 &\star \\ \m x_0 & -\m S} \preceq 0\\ 
			&\hspace{-0.6cm}\bmat{-\frac{u_{\max}^2}{\rho^2}\m S + \frac{1}{2}K_{l_1}(\cdot)& \star & \star & \star \\ \frac{1}{2}K_{l_2}(\cdot) & -\mu_0\m I+\frac{1}{2}K_{l_3}(\cdot)& \star & \star\\
				\frac{1}{\sqrt{2}}\mu_0 \m I & \m O & -\m I & \star \\
				\frac{1}{\sqrt{2}}\m Z & \frac{1}{\sqrt{2}}\mu_0 \m I & \m O & -\m I} \preceq 0 \\
			&\hspace{-0.6cm}\begin{bmatrix}
				\frac{1}{4}H_{l}(\cdot)+\mu_2-\hslash & \star \\ \frac{1}{2}(\mu_0+\mu_1) & -1
			\end{bmatrix} \preceq 0 
		\end{align}
	\end{subequations}
	where $\m \Omega = 	\m S\m A^{\top}+\m A\m S +\frac{1}{4}F_{l_1}(\cdot) +\m Z^{\top}\m B_u^{\top}+\m B_u\m Z$, and $F_{l_1},F_{l_2},G_{l},K_{l_1},K_{l_2},K_{l_3}$, and $H_{l}$ are all linear matrix-valued functions of the optimization variables given by
	\begin{subequations}~\label{equ:linforms}
		\begin{align*}
			F_{l_1}(\alpha,\m S;\tilde{\alpha},\tilde{\m S}) = \,&\tilde{\alpha}^2\m I-2\tilde{\alpha}\tilde{\m S}+\tilde{\m S}^2\nonumber-2\alpha(\tilde{\alpha}\m I-\tilde{\m S}) \\ &+(2\tilde{\alpha}\m I-\tilde{\m S})\m S-\m S\tilde{\m S}\\
			F_{l_2}(\alpha,\mu_0;\tilde{\alpha},\tilde{\mu}_0) = &(\tilde{\alpha}^2+\tilde{\mu_0}^2+2\tilde{\alpha}\tilde{\mu}_0-2\alpha(\tilde{\alpha}+\tilde{\mu}_0)\nonumber \\ &-2\mu_0(\tilde{\alpha}+\tilde{\mu}_0))\m I\\
			G_{l}(\mu_1,\m S;\tilde{\mu}_1,\tilde{\m S}) = &\,\tilde{\mu}_1^2\m I + 2\tilde{\mu}_1\tilde{\m S}+\tilde{\m S}^2-2\mu_1(\tilde{\mu}_1\m I+\tilde{\m S})\nonumber \\ &-(2\tilde{\mu}_1\m I+\tilde{\m S})\m S -\m S\tilde{\m S}
			\\
			K_{l_1}(\mu_0,\m Z;\tilde{\mu}_0,\tilde{\m Z}) = &\,\tilde{\mu}_0^2 \m I+\tilde{\m Z}^{\top}\tilde{\m Z}+\mu_0(-2\tilde{\mu}_0\m I +\tilde{\m Z}^{\top}\tilde{\m Z}) \nonumber \\
			&-\tilde{\m Z}^{\top}\m Z-\m Z^{\top}\tilde{\m Z}\\
			K_{l_2}(\m Z;\tilde{\mu}_0,\tilde{\m Z}) = &-\tilde{\mu}_0\tilde{\m Z}+\tilde{\mu}_0\m Z  \\
			K_{l_3}(\mu_0;\tilde{\mu}_0) = 	& \,(\tilde{\mu}_0^2-2\mu_0\tilde{\mu}_0)\m I\\
			H_{l}(\mu_0,\mu_1;\tilde{\mu}_0,\tilde{\mu}_1) = &\;\tilde{\mu}_0^2+\tilde{\mu}_1^2-2\tilde{\mu}_0\tilde{\mu}_1-2\mu_0(\tilde{\mu}_0-\tilde{\mu}_1)\nonumber \\ &-2\mu_1(\tilde{\mu}_1-\tilde{\mu}_0).
		\end{align*}
	\end{subequations}
	which depicts the SDP~\eqref{equ:LINF-Niko} in Theorem~\ref{th2} and $\mathcal{\m C}_t(\m S,\m Z, \alpha, \mu_0, \mu_1, \mu_2, \hslash;\tilde{\m S}, \tilde{\alpha},\tilde{\mu}_0,\tilde{\mu}_1 ) \preceq 0$ represents the block-diagonal representation of of the LMIs in~\eqref{equ:LINF-NikoP}.
\end{IEEEproof}
\section{Proof of Theorem~\ref{th:dec}}~\label{app:DLINF}
\vspace{-0.5cm}
\begin{IEEEproof}[Proof of Theorem~\ref{th:dec}]
	First BMI in~\eqref{equ:LINFDec} can be written as
	\begin{align}
		~\label{equ:LINFDec-BMI1-1}
		&\begin{bmatrix}
			\m A^{\top}\m P+\m P\m A & \m P\m B_w \\
			\m B_w^{\top}\m P & \m O
		\end{bmatrix} 
		\nonumber \\ &\quad \; + 
		\begin{bmatrix}
			\m K^{\top}\m B_u^{\top}\m P + \m P\m B_u\m K + \alpha \m P & \m O \\
			\m O & -\alpha\mu_0\m I 
		\end{bmatrix} \preceq 0.
	\end{align}
	Note that only the second term of the LHS of~\eqref{equ:LINFDec-BMI1-1} represents a BMI. Let this BMI be replaced by two convex functions  $\mc C_1(\m P,\m K;\tilde{\m P},\tilde{\m K})$ and  $\mc C_2(\alpha,\mu_0,\m P;\tilde{\alpha},\tilde{\mu}_0,\tilde{\m P})$ such that
	\begin{align}
		~\label{equ:LINFDec-BMI1-2}
		&\begin{bmatrix}
			\m A^{\top}\m P+\m P\m A & \m P\m B_w \\
			\m B_w^{\top}\m P & \m O
		\end{bmatrix} +
		\begin{bmatrix}
			\m K^{\top}\m B_u^{\top}\m P + \m P\m B_u\m K & \m O \\
			\m O & \m O 
		\end{bmatrix} \nonumber \\ &+ \begin{bmatrix}
			\alpha \m P & \m O \\
			\m O & -\alpha\mu_0\m I 
		\end{bmatrix} \preceq \begin{bmatrix}
			\m A^{\top}\m P+\m P\m A & \m P\m B_w \\
			\m B_w^{\top}\m P & \m O
		\end{bmatrix} \nonumber \\
		&\quad\;\;+\mc C_1(\m P,\m K;\tilde{\m P},\tilde{\m K}) + \mc C_2(\alpha,\mu_0,\m P;\tilde{\alpha},\tilde{\mu}_0,\tilde{\m P}).
	\end{align}
	In fact, to satisfy~\eqref{equ:LINFDec-BMI1-2}, we can leave the linear part of~\eqref{equ:LINFDec-BMI1-2} as is and require the bilinear terms to satisfy
	\begin{subequations}~\label{equ:LINFDec-BMI1-3}
		\begin{align}
			\hspace{-1cm}	\begin{bmatrix}
				\alpha \m P & \m O \\
				\m O & -\alpha\mu_0\m I 
			\end{bmatrix} &\preceq \mc C_2(\alpha,\mu_0,\m P;\tilde{\alpha},\tilde{\mu}_0,\tilde{\m P}) ~\label{equ:LINFDec-BMI1-3a} \\
			\begin{bmatrix}
				\m K^{\top}\m B_u^{\top}\m P + \m P\m B_u\m K & \m O \\
				\m O & \m O 
			\end{bmatrix} &\preceq \mc C_1(\m P,\m K;\tilde{\m P},\tilde{\m K}). ~\label{equ:LINFDec-BMI1-3b}
		\end{align}
	\end{subequations}
	Realize that~\eqref{equ:LINFDec-BMI1-3a} is similar to~\eqref{equ:LINF-BMI1-2} if $\m S$ in~\eqref{equ:LINF-BMI1-2} is replaced by $\m P$. Hence, we can use the previous result to obtain $\mc C_2(\alpha,\mu_0,\m P;\tilde{\alpha},\tilde{\mu}_0,\tilde{\m P})$, which is given as 
	\begin{align}
		\label{equ:LINFDec-BMI1-4}
		&\mc C_2(\alpha,\mu_0,\m P;\tilde{\alpha},\tilde{\mu}_0,\tilde{\m P}) = \frac{1}{4} F_{l}(\alpha,\mu_0,\m P;\tilde{\alpha},\tilde{\mu_0},\tilde{\m P}) \nonumber \\&+\begin{bmatrix}
			\frac{1}{2}(\alpha\m I+\m P) & \m O \\ \m O & \frac{1}{2}(\alpha-\mu_0)\m I
		\end{bmatrix}^{\top}\begin{bmatrix}
			\frac{1}{2}(\alpha\m I+\m P) & \m O \\ \m O & \frac{1}{2}(\alpha-\mu_0)\m I
		\end{bmatrix}
	\end{align}
	where
	\begin{align}
		\label{equ:LINFDec-BMI1-5}
		F_{l}(\alpha,\mu_0,\m P;\tilde{\alpha},\tilde{\mu}_0,\tilde{\m P}) = \begin{bmatrix}
			F_{l_1}(\alpha,\m P;\tilde{\alpha},\tilde{\m P}) & \m O \\
			\m O & F_{l_2}(\alpha,\mu_0;\tilde{\alpha},\tilde{\mu}_0)
		\end{bmatrix} 
	\end{align}
	with
	\begin{align}
		\label{equ:LINFDec-BMI1-6}
		F_{l_1}(\alpha,\m P;\tilde{\alpha},\tilde{\m P}) = \,&\tilde{\alpha}^2\m I-2\tilde{\alpha}\tilde{\m P}+\tilde{\m P}^2\nonumber-2\alpha(\tilde{\alpha}\m I-\tilde{\m P}) \\ &+(2\tilde{\alpha}\m I-\tilde{\m P})\m P-\m P\tilde{\m P}
	\end{align}
	and $F_{l_2}(\alpha,\mu_0;\tilde{\alpha},\tilde{\mu}_0)$  being equal to~\eqref{equ:LINF-BMI1-10}. Next, the bilinear terms in~\eqref{equ:LINFDec-BMI1-3b} can be expressed as
	\begin{align}
		~\label{equ:LINFDec-BMI1-7}
		\hspace{-0.4cm}\m K^{\top}\m B_u^{\top}\m P + \m P\m B_u\m K = &\,\frac{1}{2}((\m P + \m B_u \m K)^{\top}(\m P + \m B_u \m K)\nonumber\\&-(\m P - \m B_u \m K)^{\top}(\m P - \m B_u \m K)),
	\end{align}
	which is a difference of convex-concave functions. The concave part of~\eqref{equ:LINFDec-BMI1-7}, which is $-(\m P - \m B_u \m K)^{\top}(\m P - \m B_u \m K)$, can be approximated by a linear function $ G_{l}(\m P,\m K;\tilde{\m P},\tilde{\m K})$ that is calculated around $(\tilde{\m P},\tilde{\m K})$. Consequently, for all $\tilde{\m P},\tilde{\m K}$ and $\m P,\m K$, we have
	\begin{align}
		~\label{equ:LINFDec-BMI1-8}
		-(\m P - \m B_u \m K)^{\top}(\m P - \m B_u \m K) \preceq G_{l}(\m P,\m K;\tilde{\m P},\tilde{\m K})
	\end{align}
	which consequently implies
	\begin{align}
		~\label{equ:LINFDec-BMI1-9}
		\m K^{\top}\m B_u^{\top}\m P + \m P\m B_u\m K \preceq &\,\frac{1}{2}(\m P + \m B_u \m K)^{\top}(\m P + \m B_u \m K) \nonumber \\ &+  \frac{1}{2}G_{l}(\m P,\m K;\tilde{\m P},\tilde{\m K}).
	\end{align}
	Substituting $\m P = \tilde{\m P}+\Delta\m P$ and $\m K = \tilde{\m K} + \Delta \m K$ with $\Delta \m P = \m P -\tilde{\m P}$ and $\Delta \m K = \m K -\tilde{\m K}$ into the concave part of~\eqref{equ:LINFDec-BMI1-7} and removing all second-order terms that contain both $\Delta \m P$ and $\Delta \m K$ yields
	\begin{align}
		~\label{equ:LINFDec-BMI1-10}
		G_{l}(\m P,\m K;\tilde{\m P},\tilde{\m K}) = & \;\;\tilde{\m P}^2-\tilde{\m P}\m B_u \tilde{\m K}-\tilde{\m K}^{\top}\m B_u^{\top}\tilde{\m P}\nonumber \\ &+\tilde{\m K}^{\top}\m B_u^{\top}\m B_u\tilde{\m K}-\tilde{\m P}\m P-\m P\tilde{\m P}\nonumber \\ &+\tilde{\m P}\m B_u\m K+\m K^{\top}\m B_u^{\top}\tilde{\m P}+\m P\m B_u \tilde{\m K}\nonumber\\&+\tilde{\m K}^{\top}\m B_u^{\top}\m P-\tilde{\m K}^{\top}\m B_u^{\top}\m B_u\m K\nonumber\\&-\m K^{\top}\m B_u^{\top}\m B_u\tilde{\m K}.
	\end{align}
	Due to this result, $\mc C_1(\m P,\m K;\tilde{\m P},\tilde{\m K})$ can now be expressed as
	\begin{align}
		\hspace{-0.6cm}	~\label{equ:LINFDec-BMI1-11}
		\mc C_1(\m P,\m K;\tilde{\m P},\tilde{\m K}) = \begin{bmatrix}
			\frac{1}{2}(\m P + \m B_u \m K)^{\top}(\m P + \m B_u \m K) \\ + \frac{1}{2}G_{l}(\m P,\m K;\tilde{\m P},\tilde{\m K}) & \m O \\ \m O & \m O
		\end{bmatrix}.
	\end{align}
	Combining the results from and~\eqref{equ:LINFDec-BMI1-2},~\eqref{equ:LINFDec-BMI1-3},~\eqref{equ:LINFDec-BMI1-4},~\eqref{equ:LINFDec-BMI1-11} yields
	{\begin{align}
			~\label{equ:LINFDec-BMI1-12}
			&\begin{bmatrix}
				\frac{1}{2}(\alpha\m I+\m P) & \m O \\ \m O & \frac{1}{2}(\alpha-\mu_0)\m I
			\end{bmatrix}^{\top}\begin{bmatrix}
				\frac{1}{2}(\alpha\m I+\m P) & \m O \\ \m O & \frac{1}{2}(\alpha-\mu_0)\m I
			\end{bmatrix}\nonumber\\ 
			&+\begin{bmatrix}
				\m A^{\top}\m P+\m P\m A + \frac{1}{2}G_{l}(\cdot) \\
				+ \frac{1}{4} F_{l_1}(\cdot) & \m P\m B_w \\
				+ \frac{1}{2}(\m P + \m B_u \m K)^{\top}(\m P + \m B_u \m K)  \\
				\m B_w^{\top}\m P &\frac{1}{4} F_{l_2}(\cdot)
			\end{bmatrix} \preceq 0
	\end{align}}
	\noindent which is equal to the right-hand side of~\eqref{equ:LINFDec-BMI1-2}. By applying Schur complement,~\eqref{equ:LINFDec-BMI1-12} can be written as
	{\begin{align}
			\label{equ:LINFDec-BMI1-13}
			&\begin{bmatrix}
				\m A^{\top}\m P+\m P\m A \\ +\frac{1}{2}G_{l}(\cdot)  & \star & \star & \star & \star \\+\frac{1}{4} F_{l_1}(\cdot) \\ 
				\m B_w^{\top}\m P & \frac{1}{4}F_{l_2}(\cdot) & \star & \star & \star \\
				\frac{1}{2}(\alpha\m I+\m P) & \m O & -\m I & \star & \star \\
				\m O & \frac{1}{2}(\alpha-\mu_0)\m I & \m O & -\m I & \star \\
				\frac{1}{\sqrt{2}}(\m P + \m B_u \m K) & \m O & \m O & \m O & -\m I
			\end{bmatrix}  \preceq 0
	\end{align}}
	which is linear in $ \alpha,\,\mu_0,\,\m P$, and $\m K$. The last BMI in problem~\eqref{equ:LINFDec} is~\eqref{equ:LINFDec-BMI2}. In fact, the bilinear term in~\eqref{equ:LINFDec-BMI2} is similar to the one in~\eqref{equ:LINF-BMI2}, if $\m S$ in the bilinear term in~\eqref{equ:LINF-BMI2} is replaced with $\m P$. Therefore, the previous result can be used with some modifications. If $\m S$ is replaced with $\m P$ and $\m C\m S+\m D\m Z$ is replaced with $\m C+\m D\m K$ along with their respective symmetric terms, we obtain
	\begin{align}
		\label{equ:LINFDec-BMI2-1}
		\begin{bmatrix}
			\frac{1}{4}H_{l}(\mu_1,\m P;\tilde{\mu}_1,\tilde{\m P}) & \star & \star & \star\\ \m O & -\mu_2\m I & \star & \star\\ \m C +\m D \m K& \m O & -\m I & \star \\ \frac{1}{2}(\mu_1\m I - \m P) & \m O & \m O & -\m I
		\end{bmatrix} \preceq 0
	\end{align}
	that is linear in $\mu_1$ and $\m P$ where $H_{l}(\cdot)$ is defined as
	\begin{align}
		~\label{equ:LINFDec-BMI2-2}
		H_{l}(\mu_1,\m P;\tilde{\mu}_1,\tilde{\m P}) = &\,\tilde{\mu}_1^2\m I + 2\tilde{\mu}_1\tilde{\m P}+\tilde{\m P}^2-2\mu_1(\tilde{\mu}_1\m I+\tilde{\m P})\nonumber \\ &-(2\tilde{\mu}_1\m I+\tilde{\m P})\m P -\m P\tilde{\m P}.
	\end{align}
	By combining these results, the convex approximation of~\eqref{equ:LINFDec} around the point $(\tilde{\alpha},\tilde{\mu}_0,\tilde{\mu}_1,\tilde{\m P},\tilde{\m K})$ can be formulated as
	\begin{subequations}\label{equ:LINFDec-Niko}
		\begin{eqnarray}
		\min 	&& \hslash \\
		\st&& \eqref{equ:LINFDec-BMI1-13}, \eqref{equ:LINFDec-BMI2-1}\\
		&&\begin{bmatrix}
		\frac{1}{4}H_{l}(\mu_0,\mu_1;\tilde{\mu}_0,\tilde{\mu}_1)+\mu_2-\hslash & \star \\ \frac{1}{2}(\mu_0+\mu_1) & -1
		\end{bmatrix} \preceq 0 \\
		&& \{\alpha, \mu_0, \mu_1, \mu_2, \hslash\}> 0, \; \m K \in \mathcal{K}.
		\end{eqnarray}
	\end{subequations}
	This completes the proof.
\end{IEEEproof}

	\end{document}